\documentclass[aps,prx,amsmath,amssymb,amsfonts,reprint,twocolumn,floatfix]{revtex4-2}
\usepackage[T1]{fontenc}
\usepackage{times}
\usepackage{bm}
\usepackage{graphicx}
\usepackage{dcolumn}
\newcolumntype{d}[1]{D{.}{.}{#1}}
\newcolumntype{m}{>{$}c<{$}}
\allowdisplaybreaks

\newcommand*{\centt}[1]{\multicolumn{1}{c}{#1}}
\newcommand{\lbr}{\langle}
\newcommand{\rbr}{\rangle}
\newcommand{\rg}{r_\gamma}
\newcommand{\dg}{\delta^{(3)}(r_\gamma)}

\begin{document}
\preprint{Version 1.0}

\title{QED theory of scalar hyperfine interactions in HD$^+$}

\author{Tymon Kilich}
\affiliation{Faculty of Applied Physics and Mathematics, Gdańsk University of Technology, 80-233 Gdańsk, Poland}
\author{Mateusz Pańtak}
\affiliation{Faculty of Physics, University of Warsaw, Pasteura 5, 02-093 Warsaw, Poland}
\author{Krzysztof Pachucki}
\affiliation{Faculty of Physics, University of Warsaw, Pasteura 5, 02-093 Warsaw, Poland}

\date{\today}

\begin{abstract}
HD$^+$ is the simplest heteronuclear molecular ion. It has been used to determine the proton-to-electron and deuteron-to-electron mass ratios, 
to test quantum electrodynamics, and to search for new interactions. All these applications rely on an accurate knowledge of its hyperfine structure, 
for which theoretical predictions have disagreed with several precise measurements. 
Here we calculate the two dominant hyperfine interactions, 
namely the couplings of the electron spin to the proton and deuteron spins, 
including the complete nonrecoil electronic relativistic and leading radiative corrections 
of relative orders $\alpha^2$ and $\alpha^3$, respectively, evaluated in the Born–Oppenheimer approximation.
Nuclear-structure effects are suppressed by using the measured hyperfine splittings of atomic hydrogen and deuterium. 
The resulting couplings for 40 rovibrational states are about 18 times more accurate than the previous ones. 
They agree with the recent Penning-trap measurement in the ground state within the combined uncertainty of 65 Hz, i.e., 7 parts in $10^8$, 
and with a recent measurement in a rotationally excited state, whereas a discrepancy of 7.4(9) kHz persists for a vibrational overtone transition. 
Our results provide improved theoretical input for future determinations of the proton and deuteron masses from HD$+$ spectroscopy.
\end{abstract}
\maketitle

\section{Introduction}\label{sec:intro}
The hydrogen molecular ions are the simplest molecules: two nuclei bound by a single electron.
Such three-body molecular ions can be both accurately measured \cite{patra:20,alighanbari:20,kortunov:21,alighanbari:23,kliukin:26, konig:26, bressel:12,biesheuvel:16,carrington:91} 
and accurately calculated from quantum electrodynamics (QED) \cite{korobov:17,korobov:21}.  
For this reason, HD$^+$ is currently used for the most accurate determination of 
the proton and deuteron masses in units of the electron mass \cite{codata:22,karr:23}. HD$^+$ spectra have also been used 
in searches for long-range interactions beyond the Standard Model (BSM) \cite{germann:21,delaunay:23}. 

Every rovibrational level of HD$^+$ is split by the magnetic interactions between the spins of the electron, the proton, and the deuteron, and the rotation of the molecule.
This hyperfine structure (hfs) has to be accurately accounted for in order to obtain the spin-averaged transition frequencies 
from which the mass ratios are determined \cite{karr:23}. In addition,
the hyperfine intervals themselves are measured with high accuracy and provide an independent test of the theory.
However, significant discrepancies between several measurements and theoretical predictions have been reported
for the HD$^+$ hyperfine splitting \cite{haidar:22, karr:23, konig:26, kliukin:26}. 
This is surprising, because in another three-body system, namely $^3$He, theoretical predictions for
the hyperfine splitting of the $2^3S_1$ state are in very good agreement with the measured values, down to the uncertainty level of 44 Hz.
This agreement has been achieved by a complete calculation of the radiative $\alpha\,(Z\,\alpha)^2\,E_F$ correction \cite{patkos:23prl},
where $E_F$ is the leading (Fermi) hyperfine energy, $\alpha$ is the fine-structure constant, and $Z$ is the nuclear charge number. 
We show in this work that the $\alpha\,(Z\,\alpha)^2\,E_F$ correction to the electron--nucleus spin interaction indeed explains
the previously reported  discrepancies for the hyperfine splitting in the ground $(v,L) = (0,0)$ \cite{konig:26} and in the rotational $(0,3)$  \cite{kliukin:26} levels
of HD$^+$, while the hyperfine splitting in the $(9,3)$ \cite{patra:20}  level remains discrepant.

The hyperfine structure of an arbitrary rovibrational level of HD$^+$ can be described by
an effective Hamiltonian, which consists of all possible couplings between the three spins $\vec s_e, \vec s_p, \vec s_d$
and the angular momentum $\vec L$,
\begin{align}
H_\mathrm{eff} =&\
A_{ep}\,\vec s_e\cdot\vec s_p +  A_{ed}\,\vec s_e\cdot\vec s_d + A_{el}\,\vec s_e\cdot\vec L
\nonumber \\ &
+ A_{pl}\,\vec s_p\cdot\vec L + A_{dl}\,\vec s_d\cdot\vec L + A_{pd}\,\vec s_p\cdot\vec s_d +\ldots\,, \label{eq:1}
\end{align}
where $s_e=1/2$, $s_p=1/2$, and $s_d=1$, and the dots denote terms involving four 
or more powers of spin and angular momentum operators.
We concentrate here only on the first two coefficients, $A_{ep}$ and $A_{ed}$ ($E_4$ and $E_5$ in Korobov's notation \cite{bakalov:06}).
These two terms are the scalar  interactions of the electron spin with the nuclear spins, at the leading order
proportional to the electron density at the corresponding nucleus, and are the largest terms in $H_\mathrm{eff}$.
They have, in principle, already been calculated \cite{bakalov:06,korobov:16,korobov:20,karr:20}, 
but these calculations missed the $(Z\,\alpha)^2\,E_F$ second-order spin--orbit and $\alpha\,(Z\,\alpha)^2\,E_F$ state-dependent corrections.

The calculation of $A_{ep}$ and $A_{ed}$ entirely from first principles would be severely 
limited by the poorly known nuclear-structure effects.
Instead, we rescale the measured hyperfine splittings of atomic hydrogen and deuterium 
by the ratio of the electron densities at the nucleus in the molecular ion and in the atom.
All contributions proportional to this density, including the nuclear-structure ones, are then accounted for by the experimental atomic values, 
and only the state-dependent contribution is calculated using nonrelativistic quantum electrodynamics (NRQED) and an expansion in $\alpha$.
It is similar to previous calculations by Karr {\em et al.}~\cite{karr:20}, where the nuclear-structure correction was extracted 
from the atomic hyperfine splitting and was then used in HD$^+$. In our approach, we do not need to consider the contact QED corrections separately,
because they are automatically included by rescaling the atomic hyperfine splitting.
As a result, only the relativistic $(Z\,\alpha)^2\,E_F$ and radiative $\alpha\,(Z\,\alpha)^2\,E_F$ state-dependent
corrections remain to be calculated, while higher-order terms are neglected, because they are too small to matter at the present level of precision. 
We derive these two corrections in closed form and evaluate them in the Born--Oppenheimer (BO) approximation,
verifying for each of them that the limit of separated atoms reproduces the known hydrogenic result.

The paper is organized as follows. In Sec.~\ref{sec:overview} we give an overview of the method 
and in Sec.~\ref{sec:results} we present the main results and compare them with experiments.
The notation and the BO approximation are introduced in Sec.~\ref{sec:bo}, the relativistic and QED corrections to the BO energy in Sec.~\ref{sec:energy},
and the expansion of the hyperfine couplings in Sec.~\ref{sec:Aeg}. The theory of the hfs in hydrogen-like atoms is summarized in Sec.~\ref{sec:hydrogen}.
The relativistic correction is derived in Sec.~\ref{sec:A6}. The final formulas for the radiative correction are presented in Sec.~\ref{sec:A7}, 
and their derivation in Appendix~\ref{app:A7}. Section~\ref{sec:num} describes the numerical calculations, Sec.~\ref{sec:karr} compares our results with previous calculations, 
and Sec.~\ref{sec:summary} contains the summary and outlook.
Technical details of the numerical calculations and the tabulated potential curves are collected in Appendix~\ref{app:num}.

\section{Overview of the method}\label{sec:overview}
Because we aim at an uncertainty much smaller than the nuclear-structure effects, 
which are difficult to calculate, the scalar couplings for a rovibrational state $(v,L)$ are obtained as 
\begin{align}
A_{e\gamma}(v,L) =&\ \bar A_{e\gamma}(v,L) + A^{(6)}_{e\gamma}(v,L) + A^{(7)}_{e\gamma}(v,L) + \ldots\,, \label{eq:2}
\end{align}
a sum of three terms, where $\gamma=p,d$ denotes the nucleus and the dots stand for the neglected higher-order terms. The first term,
\begin{align}
\bar A_{e\gamma}(v,L) = \frac{\langle\pi\,\delta^{(3)}(r_{e\gamma})\rangle_{vL}}{\langle\pi\,\delta^{(3)}(r)\rangle_{\gamma}}\,A_{\gamma}^\mathrm{exp}\,, \label{eq:3}
\end{align}
is the measured hyperfine constant $A_{\gamma}^\mathrm{exp}$ of the hydrogen or deuterium atom, 
multiplied by the ratio of the electron densities at the nucleus in HD$^+$ and in the atom.
The density in HD$^+$ is calculated with the full three-body (nonadiabatic) wave function \cite{pantak:26}.
This term includes the leading Fermi interaction together with those higher-order corrections that are proportional to the electron density at the nucleus,
in particular the nuclear-structure effects, which cannot presently be calculated with sufficient accuracy.
The second and third terms in Eq.~(\ref{eq:2}) are the remaining corrections of orders $(Z\,\alpha)^2\,E_F$ and $\alpha\,(Z\,\alpha)^2\,E_F$ 
that are beyond  this density scaling. They depend on the molecular state and are the subject of this work.
We derive them within NRQED and express them in terms of functions of the internuclear distance $R$, 
which are calculated in the BO approximation and then averaged with the nuclear wave function; 
see Eq.~(\ref{eq:42}) for the exact definition. 
The complete calculation of these two corrections is the source of the difference from the previous results \cite{karr:20}; see Sec.~\ref{sec:karr} for details.
The magnitudes of the individual terms for the ground state are as follows.
For the electron--proton coupling, $\bar A_{ep}\approx 925$~MHz, $A^{(6)}_{ep}\approx -7.7$~kHz, and $A^{(7)}_{ep}\approx -0.54$~kHz.
The uncertainty due to the uncalculated higher-order terms is estimated to be 51~Hz for $A_{ep}$ and 2~Hz for $A_{ed}$, 
which is comparable to or smaller than the uncertainty of the most accurate measurement \cite{konig:26}.

\section{Results and comparison with experiments} \label{sec:results}
\begin{figure}[tb]
\includegraphics[width=\columnwidth]{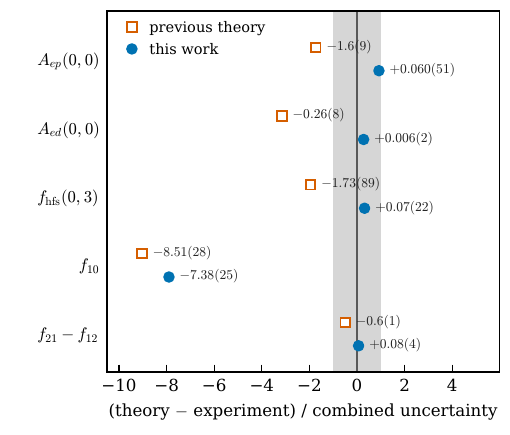}
\caption{\label{fig:comparison} Deviations of the theoretical predictions from the measured values, divided by the combined theoretical and experimental uncertainty: 
previous theory (open squares) and this work (full circles). The numbers next to the points are the deviations in kHz, with the theoretical uncertainty in parentheses,
and the shaded band is the range of $\pm1$ combined standard deviation.
From top to bottom: the scalar couplings $A_{ep}$ and $A_{ed}$ in the ground state $(v,L)=(0,0)$, Table~\ref{tab:II}; 
the hyperfine interval $f_\mathrm{hfs}$ in the $(v,L)=(0,3)$ state, Table~\ref{tab:III}; 
the difference $f_{10}$ of two hyperfine components of the $(0,3)\to(9,3)$ transition, Table~\ref{tab:IV};
and the difference $f_{21}-f_{12}$ of two hyperfine components of the $(0,0)\to(5,1)$ transition, Table~\ref{tab:V},
where the two experimental uncertainties are added in quadrature.}
\end{figure}
Our results for the scalar couplings $A_{ep}$ and $A_{ed}$ in the rovibrational states with $0\leq v\leq 9$ and $0\leq L\leq 3$ are presented in Table~\ref{tab:I},
and their comparison with experiments and with the previous theoretical predictions is summarized in Fig.~\ref{fig:comparison}.
\begin{table*}[htb]
\caption{\label{tab:I} Electron--proton and electron--deuteron scalar couplings $A_{ep}(v,L)$ and $A_{ed}(v,L)$ (in kHz),
from the last columns of Tables~\ref{tab:X} and \ref{tab:XI}, arranged by the vibrational ($v$) and rotational ($L$) quantum numbers. 
The uncertainties due to the unknown higher-order terms of Eq.~(\ref{eq:42}) are estimated as 
$2\,g_\gamma\,m/M_\gamma\,|A^{(6)}_{e\gamma}|+\alpha\,|A^{(7)}_{e\gamma}|$, 
with $A^{(6)}_{e\gamma}$ and $A^{(7)}_{e\gamma}$  from Tables~\ref{tab:X} and \ref{tab:XI}.}
\footnotesize
\begin{ruledtabular}
\begin{tabular}{m*{4}{d{6.7}}*{4}{d{6.6}}}
& \multicolumn{4}{c}{$A_{ep}$ [kHz]} & \multicolumn{4}{c}{$A_{ed}$ [kHz]}\\
\centt{$v$} & \centt{$L=0$} & \centt{$L=1$} & \centt{$L=2$} & \centt{$L=3$} & \centt{$L=0$} & \centt{$L=1$} & \centt{$L=2$} & \centt{$L=3$}\\ \hline
0 & 925395.818(51) & 924569.375(51) & 922925.432(51) & 920481.620(51) & 142287.827(2) & 142160.941(2) & 141908.544(2) & 141533.344(2)\\
1 & 904148.672(49) & 903368.031(49) & 901815.307(49) & 899507.360(49) & 139030.350(2) & 138910.518(2) & 138672.169(2) & 138317.893(2)\\
2 & 884290.280(47) & 883553.314(47) & 882087.570(47) & 879909.182(47) & 135987.377(2) & 135874.276(2) & 135649.330(2) & 135315.022(1)\\
3 & 865742.407(45) & 865047.215(45) & 863664.670(45) & 861610.217(45) & 133147.139(1) & 133040.480(1) & 132828.369(1) & 132513.178(1)\\
4 & 848434.919(43) & 847779.812(43) & 846477.109(43) & 844541.599(43) & 130499.186(1) & 130398.718(1) & 130198.937(1) & 129902.117(1)\\
5 & 832304.973(41) & 831688.462(41) & 830462.636(41) & 828641.664(41) & 128034.294(1) & 127939.797(1) & 127751.909(1) & 127472.809(1)\\
6 & 817296.274(39) & 816717.053(39) & 815565.509(39) & 813855.222(39) & 125744.392(1) & 125655.677(1) & 125479.307(1) & 125217.373(1)\\
7 & 803358.357(37) & 802815.294(37) & 801735.781(37) & 800132.840(37) & 123622.516(1) & 123539.425(1) & 123374.262(1) & 123129.031(1)\\
8 & 790445.837(35) & 789937.960(35) & 788928.551(35) & 787430.098(35) & 121662.795(1) & 121585.203(1) & 121430.998(1) & 121202.105(1)\\
9 & 778517.552(33) & 778044.037(33) & 777103.096(33) & 775706.709(33) & 119860.481(1) & 119788.297(1) & 119644.869(1) & 119432.050(1)\\
\end{tabular}
\end{ruledtabular}
\end{table*}
The uncertainties due to the unknown higher-order terms of Eq.~(\ref{eq:42}) are estimated as follows.
The uncertainty due to the Born--Oppenheimer approximation and the neglect of the $(Z\,\alpha)^2\,m/M_\gamma\,E_F$ correction
is estimated by $2\,g_\gamma\,m/M_\gamma\,|A^{(6)}_{e\gamma}|$, where the enhancement factor $g_\gamma$ comes from the second order
hyperfine interaction, and a single power of $m/M$ comes from the so-called nonadiabatic perturbation theory  \cite{napt}.
The uncertainty due to the neglect of $A^{(8)}_{e\gamma}$
is estimated by $\alpha\,|A^{(7)}_{e\gamma}|$, because we do not expect any logarithmic enhancements at this order.
The final uncertainty is the sum of these two. Analogously, for the difference of the couplings in two states we assume the uncertainty of
$2\,g_\gamma\,m/M_\gamma\,|A^{(6)}_{e\gamma}(v,L) - A^{(6)}_{e\gamma}(v',L')|+\alpha\,|A^{(7)}_{e\gamma}(v,L) - A^{(7)}_{e\gamma}(v',L')|$.
We should nevertheless mention that estimates of terms that have not yet been calculated are not necessarily reliable.
One example is the work of Ref.~\cite{karr:20}, and another is Ref.~\cite{zhong:26}, where the nonadiabatic effects
appear to be 10 times larger than previously estimated \cite{korobov:17,korobov:21}, although this concerns spin-independent transitions.
 
In comparison with the previous calculations by Karr {\em et al.} \cite{karr:20}, our results are 18 times more accurate and differ 
by up to about three of their standard deviations. The source of the difference is the complete (in the BO approximation)
calculation of the $(Z\,\alpha)^2\,E_F$ and $\alpha\,(Z\,\alpha)^2\,E_F$ corrections. 
We have checked the correctness of our results by verifying the large-$R$ asymptotics of
${\cal A}^{(7)}_{e\gamma}(R)$, as well as the asymptotics of all other potential curves against the known hydrogenic results.  

In Table~\ref{tab:II} we compare our results for the scalar hyperfine interactions in the ground rovibrational state $(v,L) = (0,0)$ 
 with the very recent measurement by K\"onig {\em et al.} \cite{konig:26} and find good agreement, well within the uncertainties.
\begin{table}[htb]
\caption{\label{tab:II} The scalar spin couplings $A_{ep}$ and $A_{ed}$ of HD$^+$ in the ground rovibrational state $(v,L)=(0,0)$, in kHz:
the recent measurement by K\"onig \emph{et al.} \cite{konig:26}, with the statistical and systematic uncertainties in the first and second parentheses,
this work (Table~\ref{tab:I}), and the previous calculation of Karr \emph{et al.} \cite{karr:20}.}
\begin{ruledtabular}
\begin{tabular}{ld{7.11}d{7.10}}
 & \centt{$A_{ep}(0,0)$} & \centt{$A_{ed}(0,0)$} \\ \hline
 Experiment \cite{konig:26} & 925\,395.758(39)(11) & 142\,287.821(20)(8) \\
This work & 925\,395.818(51) & 142\,287.827(2) \\
Karr \emph{et al.}~\cite{karr:20} & 925\,394.2(9) & 142\,287.56(8) \\
\end{tabular}
\end{ruledtabular}
\end{table}

In Table~\ref{tab:III} we compare our result for the hyperfine transition $(F,S,J)=(0,1,4)\rightarrow(1,2,5)$ 
in the $(v,L)=(0,3)$ rovibrational state of HD$^+$ with the very recent measurement \cite{kliukin:26} 
and with previous theoretical predictions from Refs.~\cite{korobov:20,haidar:22,kliukin:26}. 
Our result is based on the scalar couplings from Table~\ref{tab:I} and the remaining spin--orbit and tensor spin couplings from Refs.~\cite{korobov:20,haidar:22};
its uncertainty is dominated by the spin--orbit coupling $A_{el}$. It is in very good agreement with the measured value of Ref.~\cite{kliukin:26}.
\begin{table}[htb]
\caption{\label{tab:III} The total hyperfine interval $f_\mathrm{hfs}$ between the
$(F,S,J)=(0,1,4)$ and $(1,2,5)$ levels of HD$^+$ in the $(v,L)=(0,3)$ state, in kHz.
The measurement is from Ref.~\cite{kliukin:26}.
Our result is based on the scalar couplings from Table~\ref{tab:I} and all other couplings from Refs.~\cite{korobov:20,haidar:22}.
The previous theoretical predictions are from Ref.~\cite{kliukin:26} and are based on the calculations of Refs.~\cite{korobov:20,haidar:22}.}
\begin{ruledtabular}
\begin{tabular}{ld{7.7}}
 & \centt{$f_\mathrm{hfs}$} \\ \hline
Experiment~\cite{kliukin:26} & 1\,050\,804.511(11) \\
This work & 1\,050\,804.58(22) \\
Previous theory~\cite{kliukin:26} & 1\,050\,802.78(89) \\
\end{tabular}
\end{ruledtabular}
\end{table}

In Table~\ref{tab:IV} we present our result for the so-called $f_{10} = \nu_{F=0} - \nu_{F=1}$ hyperfine interval,
which is the difference between two transition frequencies \cite{patra:20}.
Using the notation $(v,L; F,S,J)$ to identify a hyperfine level,
the first ($F=0$) transition is $(0,3; 0,1,4)\rightarrow(9,3; 0,1,4)$,
and the second ($F=1$) transition is $ (0,3; 1,2,5)\rightarrow(9,3;1,2,5)$. 
Our calculation of $f_{10}$ is based on $A_{ep}$ and $A_{ed}$ from Table~\ref{tab:I} and the remaining spin--orbit and tensor spin couplings from Refs.~\cite{korobov:20,haidar:22}.
The theoretical uncertainty is again dominated by the electron spin--orbit coupling $A_{el}$,
and we observe a significant discrepancy of $7.4(9)$~kHz with the measurement \cite{patra:20}.
Since we observe very good agreement for the hfs of the $(v,L) = (0,3)$ state in Table~\ref{tab:III},
the discrepancy most likely comes from the $(v,L) = (9,3)$ state, although this comparison alone does not uniquely identify its source. 
On the theoretical side, all hyperfine coefficients
should be verified by nonadiabatic calculations at the $\alpha^2\,E_F$ order. 
On the experimental side, a hyperfine transition analogous to that of Ref.~\cite{kliukin:26} 
should be measured within the $(v,L) = (9,3)$ state.
\begin{table}[htb]
\caption{\label{tab:IV} The hyperfine interval $f_{10}$, which is the difference between two measured 
 $(0,3)\to(9,3)$ transition frequencies in HD$^+$, in kHz; see the text for details.
The measurement is from Ref.~\cite{patra:20}. We update the prediction of
Ref.~\cite{haidar:22} with the scalar couplings from Table~\ref{tab:I}.}
\begin{ruledtabular}
\begin{tabular}{ld{6.6}}
 & \centt{$f_{10}$} \\ \hline
Experiment~\cite{patra:20} & 178\,254.4(9) \\
This work & 178\,247.02(25) \\
Haidar \emph{et al.}~\cite{haidar:22} & 178\,245.89(28) \\
\end{tabular}
\end{ruledtabular}
\end{table}

In Table~\ref{tab:V} we compare our prediction for the difference $f_{21}-f_{12}$
in the $(v,L)=(0,0)\to(5,1)$ transitions with the measurement of Alighanbari \emph{et al.}~\cite{alighanbari:23}.
The lines $f_{12}$ and $f_{21}$ correspond to the transitions $(G_1,G_2,F)=(1,2,2)\to(1,2,1)$ and $(0,1,1)\to(0,1,2)$, respectively
(in the notation of Ref.~\cite{alighanbari:23}, $\vec G_1=\vec s_e+\vec s_p$, $\vec G_2=\vec G_1+\vec s_d$, and $\vec F = \vec G_2+\vec L$;
this is the same coupling scheme as in Ref.~\cite{patra:20}, only the symbols are different).
Our result is based on $A_{ep}$ and $A_{ed}$ from Table~\ref{tab:I} and the remaining couplings from Refs.~\cite{haidar:22,alighanbari:23}.
It is in slightly better agreement with the measurement than the previous theoretical prediction, 
but the experimental uncertainty is too large to draw a definite conclusion.
\begin{table}[htb]
\caption{\label{tab:V} The difference $f_{21}-f_{12}$ of the $(v,L)=(0,0)\to(5,1)$ transitions in HD$^+$, in kHz;
see the text for details. The experimental value and the previous theoretical prediction are from Eq.~(8) of Ref.~\cite{alighanbari:23}.}
\begin{ruledtabular}
\begin{tabular}{ld{6.7}}
 & \centt{$f_{21}-f_{12}$} \\ \hline
Experiment~\cite{alighanbari:23} & 126\,092.6(1.2) \\
This work & 126\,092.68(4) \\
Previous theory~\cite{alighanbari:23} & 126\,092.0(1) \\
\end{tabular}
\end{ruledtabular}
\end{table}

Although the present calculation resolves the discrepancies in the ground and rotationally excited hyperfine intervals discussed above, some disagreement remains for other rotational transitions, in particular the fundamental rotational transition considered in Ref. \cite{haidar:22}. Updating the scalar couplings calculated here does not significantly change this comparison, indicating that further investigation of the remaining hyperfine coefficients is necessary.
For instance, the difference \(f_{20}-f_{12}\) remains discrepant by approximately 1 kHz, 
much larger than its experimental uncertainty. Updating the scalar couplings calculated here changes this prediction by only about 20 Hz.

\section{Notation and Born--Oppenheimer approximation}\label{sec:bo}
To present the calculations, we first introduce the notation.
The $A$ coefficients in Eq.~(\ref{eq:1}) are expanded in the fine-structure constant $\alpha$.
In the leading $(Z\,\alpha)^4$ order
\begin{align}
A^{(4)}_{e\gamma} =&\
\frac{4}{3}\,\frac{Z\,\alpha}{m\,M_\gamma}\,g_\gamma\,\langle\pi\,\delta^{(3)}(r_{e\gamma})\rangle
\nonumber \\ =&\
(Z\,\alpha)^4\,\frac{m^2}{M_\gamma}\,g_\gamma\,{\cal A}^{(4)}_{e\gamma}\,, \label{eq:4}
\end{align}
where $\gamma = p,d$, $m\equiv m_e$ is the electron mass, $M_\gamma$ is the nuclear mass ($M_p\equiv m_p$, $M_d\equiv m_d$), and we have introduced the calligraphic symbol ${\cal A}$ to denote a dimensionless quantity. 
The nuclear ``natural'' $g$ factor is defined through the nuclear magnetic moment $\mu$,
\begin{align}
\mu =&\ \frac{Z\,|e|}{2\,M}\,g\,I\,, \label{eq:5}
\end{align}
where $I$ is the nuclear spin, $M$ is the nuclear mass, $Z$ is the nuclear charge in units of the elementary charge,
and $e = -|e|$ is the electron charge. 
The magnetic moments of nuclei are usually given in units of the nuclear magneton $\mu_\mathrm{N}$,
\begin{align}
\mu_\mathrm{N} =&\ \frac{|e|}{2\,m_p}\,. \label{eq:6}
\end{align}
Thus, the nuclear ``natural'' $g$ factors of the proton and deuteron are 
\begin{align}
  g_p &= \frac{\mu_p}{\mu_\mathrm{N}\,I_p}\,, \label{eq:7}                 \\
  g_d &= \frac{\mu_d}{\mu_\mathrm{N}\,I_d}\,\frac{m_d}{m_p}\,, \label{eq:8}
\end{align}
where $I_p =1/2$ and $I_d=1$.
The relation of the atomic constant $A_\gamma$ to the ground-state hyperfine splitting is 
$A_p =  \nu_\mathrm{hfs}(\mathrm{H})$ for hydrogen, and
$A_d = \frac{2}{3}\,\nu_\mathrm{hfs}(\mathrm{D})$ for deuterium.
Finally, these hyperfine splittings, magnetic moments, $g$ factors, mass ratios, and 
fundamental constants used in this work are presented in Table~\ref{tab:VI}.  

\begin{table}[htb]
\caption{\label{tab:VI} Physical constants \cite{codata:22} and experimental hyperfine splittings used in this work \cite{essen:71, essen:73, winelandt:72}.  
Uncertainties in the last digits are given in parentheses.}
\begin{ruledtabular}
\begin{tabular}{lll}
Quantity & Value & Unit \\ \hline
$m_p/m_e$                            & $1836.152\,673\,426(32)$               &                 \\
$m_d/m_e$                            & $3670.482\,967\,655(63)$               &                 \\
$\mu_p$                              & $2.792\,847\,344\,63(82)$              & $\mu_\mathrm{N}$         \\
$\mu_d$                              & $0.857\,438\,2335(22)$                 & $\mu_\mathrm{N}$         \\
$g_p$                                & $5.585\,694\,6893(16)$                 &                 \\
$g_d$                                & $1.714\,025\,4606(44)$                 &                 \\
$\alpha^{-1}$                        & $137.035\,999\,177(21)$                &                 \\
$R_\infty c$                         & $3.289\,841\,960\,2500(36)\times10^{15}$ & Hz            \\
$\nu^{\rm exp}_{\rm hfs}({\rm H})$   & $1\,420\,405\,751.766(3)$              & Hz              \\
$\nu^{\rm exp}_{\rm hfs}({\rm D})$   & $327\,384\,352.522(2)$                 & Hz              \\
\end{tabular}
\end{ruledtabular}
\end{table}

We derive formulas and perform calculations in the BO approximation.
The total wave function of HD$^+$ is represented in the BO approximation as 
\begin{align}
\phi_\mathrm{BO}(\vec r_A,\vec r_B)=&\
\phi(r_A, r_B, R)\,\frac{\chi_{vL}(R)}{R}\,Y_{LM}\Big(\frac{\vec R}{R}\Big)\,, \label{eq:9}
\end{align}
 where $R = |\vec R_A-\vec R_B|$, $r_A = |\vec r-\vec R_A|$, and  $r_B = |\vec r-\vec R_B|$.
The nuclei $A$ and $B$ are the proton and the deuteron, and thus they will also be denoted by $\gamma=p,d$,
and the distance between the electron and the nucleus by $r_\gamma$.
We will use the equivalent notation for the electron--nucleus distance,
$r_{e\gamma}$, when full three-body expectation values are considered. 
The {\em electronic} wave function $\phi$ satisfies
\begin{align}
[{\cal H}-{\cal E}(R)]|\phi\rangle = 0\,, \label{eq:10}
\end{align}
with the clamped-nuclei Hamiltonian~\cite{fernandez:21}
 \begin{align}
{\cal H} = \frac{p^2}{2} -\frac{1}{r_A} -\frac{1}{r_B}\,, \label{eq:11}
\end{align}
where calligraphic symbols denote dimensionless quantities.
We define the physical nonrelativistic energy as 
\begin{align}
{\cal E}^{(2)}(R) =&\ {\cal E}(R) + \frac{1}{R}\,, \label{eq:12}
\end{align}
with the large-$R$ limit
\begin{align}
\lim_{R\rightarrow\infty} {\cal E}^{(2)}(R) =&\ -\frac{1}{2}\,. \label{eq:13}
\end{align}
The leading contact hyperfine interaction in Eq.~(\ref{eq:4}) is proportional to 
\begin{align}
D_A(R) = 2\,\pi\,\langle\phi|\delta^{(3)}(r_A)|\phi\rangle\,, \label{eq:14}
\end{align}
with the large-$R$ limit
\begin{align}
\lim_{R\rightarrow\infty} D_A(R) =&\ 1\,, \label{eq:15}
\end{align}
so that ${\cal A}^{(4)}_{e\gamma}(R) = \frac{4}{3}\,\langle\pi\,\delta^{(3)}(r_{\gamma})\rangle = \frac{2}{3}\,D_\gamma(R)$ in the BO approximation.
The radial equation for $\chi_{vL}(R)$ in Eq.~(\ref{eq:9}) is
\begin{align}
({\cal H}_\mathrm{BO} - {\cal E}_{vL})\,\chi_{vL} = 0\,, \label{eq:16}
\end{align}
where
\begin{align}
{\cal H}_\mathrm{BO} = -\frac{1}{2\,\mu_\mathrm{n}}\,\frac{\partial^2}{\partial R^2} + \frac{L\,(L+1)}{2\,\mu_\mathrm{n}\,R^2} + \tilde{\cal E}^{(2)}(R)\,, \label{eq:17}
\end{align}
and where
\begin{align}
\tilde{\cal E}^{(2)}(R) = {\cal E}^{(2)}(R) - {\cal E}^{(2)}(\infty)\,, \label{eq:18}
\end{align}
and $\mu_\mathrm{n}$ is the reduced mass of the proton and deuteron in units of the electron mass,
\begin{align}
\mu_\mathrm{n} = \frac{m_p\,m_d}{m\,(m_p+m_d)}\,. \label{eq:19}
\end{align}
Finally, the radial wave function $\chi_{vL}$ satisfies the boundary condition $\chi_{vL}(0) = 0$ and the normalization condition
\begin{align}
\int_0^\infty dR\,\chi^2_{vL}(R) = 1\,. \label{eq:20}
\end{align}

\section{Relativistic and QED corrections to the energy}\label{sec:energy}
The total BO energy of HD$^+$ is expanded in powers of $\alpha$,
\begin{align}
E(R) =&\ m\,\alpha^2\Big[{\cal E}^{(2)}(R) + \alpha^2\,{\cal E}^{(4)}(R) + \frac{\alpha^3}{\pi}\,{\cal E}^{(5)}(R)+\ldots\Big]\,, \label{eq:21}
\end{align}
where ${\cal E}^{(2)}(R)$ is defined in Eq.~(\ref{eq:12}),
the leading relativistic correction~\cite{tsogbayar:06} is
\begin{align}
{\cal E}^{(4)}(R)=&\ \langle {\cal H}^{(4)}_A \rangle\,, \label{eq:22}
\end{align}
where
\begin{align}
{\cal H}^{(4)}_A = &\ -\frac{p^4}{8}+\frac{\pi}{2}\,[\delta^{(3)}(r_A) + \delta^{(3)}(r_B)]\,. \label{eq:23}
\end{align}
There is also another effective operator, $H_B^{(4)}$, of the same order in $Z\,\alpha$; see Eq.~(\ref{eq:57}).
However, its expectation value contributes only to $A_{el}$ in Eq.~(\ref{eq:1}).  
The large-$R$ asymptotics of ${\cal E}^{(4)}$ is
\begin{align}
\lim_{R\rightarrow\infty} {\cal E}^{(4)}(R) =&\ -\frac{1}{8}\,. \label{eq:24}
\end{align}

The next term of the $\alpha$ expansion in Eq.~(\ref{eq:21}) is the leading QED correction to the energy~\cite{bishop:81,bukowski:92,korobov:18bo},
\begin{align}
{\cal E}^{(5)}(R) =&\
\bigg(\frac{4}{3}\ln(\alpha^{-2})+\frac{10}{9}-\frac{4}{15} -\frac{4}{3}\,\ln k_0(R)\bigg)
\nonumber \\ &\times
\langle\pi[\delta^{(3)}(r_A) + \delta^{(3)}(r_B)]\rangle\,, \label{eq:25}
\end{align}
where
\begin{align}
\ln k_0 =&\ \frac{\langle\vec p({\cal H}-{\cal E})\,\ln[2\,({\cal H}-{\cal E})]\,\vec p\rangle}{\langle\vec p({\cal H}-{\cal E})\vec p\rangle}
\nonumber \\ =&\
\frac{\langle\vec p({\cal H}-{\cal E})\,\ln[2\,({\cal H}-{\cal E})]\,\vec p\rangle}{2\,\pi\,\langle\delta^{(3)}(r_A) + \delta^{(3)}(r_B)\rangle}\,. \label{eq:26}
\end{align}
For the numerical calculation of $\ln k_0$ we use the integral representation
\begin{align}
\ln k_0 =&\ \frac{1}{D}\,\int_0^1 dt\,\frac{f(t)-f_0-f_2\,t^2}{t^3}\,, \label{eq:27}
\end{align}
where
\begin{align}
D =&\ 2\,\pi\,\langle\delta^{(3)}(r_A) + \delta^{(3)}(r_B)\rangle = D_A + D_B\,, \label{eq:28}                 \\
f(t) =&\  \bigg\langle\vec p\frac{k}{k+{\cal H}-{\cal E}}\,\vec p\bigg\rangle\;\mbox{\rm with}\; t = \frac{1}{\sqrt{1+2\,k}}\,. \label{eq:29}
\end{align}
In the numerical calculation it is convenient to use
the small-$t$ expansion of $f(t)$ to verify the numerical precision in this difficult region,
\begin{align}
f(t) =&\ f_0 + f_2\,t^2 + f_3\,t^3 + f_4\,t^4\,\ln(t) + f_5\,t^4 + O(t^5)\,, \label{eq:30}
\end{align}
where
\begin{align}
f_0 =&\ \langle\vec p^{\,2}\rangle\,, \label{eq:31}                 \\
f_2 =&\ -2\,D\,, \label{eq:32}                  \\
f_3 =&\ 8\,D\,, \label{eq:33}                  \\
f_4 =&\ 16\,D\,, \label{eq:34}                 \\
f_5 =&\ 4\,\bigg\langle\bigg(\frac{\vec r_A}{r_A^3} + \frac{\vec r_B}{r_B^3}\biggr)^2\bigg\rangle + (16\,\ln2 + 6)\,D\,, \label{eq:35}
\end{align}
where the expectation value $\langle r_\gamma^{-4}\rangle$ is defined in Eq.~(\ref{eq:77}).
We verify the numerical calculations against the known large-$R$ asymptotics, which are
\begin{align}
\lim_{R\rightarrow\infty} f_5(R) =&\ 64\,\ln 2-20\,, \label{eq:36}
\\
\lim_{R\rightarrow\infty} \ln k_0(R)  =&\  2.984\,128\,555\,765\,497\,610\,, \label{eq:37}
\\
\lim_{R\rightarrow\infty} {\cal E}^{(5)}(R) =&\ 9.986\,256\,127\,493\,285\,. \label{eq:38}
\end{align} 

\section{Expansion of the hyperfine couplings in $\alpha$}\label{sec:Aeg}
We now turn to the hyperfine constants.
The coupling $A_{e\gamma}$ can be expanded in powers of $\alpha$ with $Z=1$,
\begin{align}
A_{e\gamma} =&\ A^{(4)}_{e\gamma} + A^{(5)}_{e\gamma} + A^{(6)}_{e\gamma} + A^{(7)}_{e\gamma} +\ldots
\nonumber \\ =&\
\alpha^4\,\frac{m^2}{M_\gamma}\,g_\gamma\,\big[
{\cal A}^{(4)}_{e\gamma} + \alpha\,{\cal A}^{(5)}_{e\gamma}
+ \alpha^2\,{\cal A}^{(6)}_{e\gamma} + \frac{\alpha^3}{\pi}\,{\cal A}^{(7)}_{e\gamma}\big]\,. \label{eq:39}
\end{align} 
Because the nuclear-structure contribution is not well known, we calculate $A_{e\gamma}$ as 
\begin{align}
A_{e\gamma} =&\
 \frac{\langle\pi\,\delta^{(3)}(r_{e\gamma})\rangle}{\langle\pi\,\delta^{(3)}(r)\rangle_{\gamma}}\,A_{\gamma}^\mathrm{exp}
 +\bigg[A_{e\gamma} -\frac{\langle\pi\,\delta^{(3)}(r_{e\gamma})\rangle}{\langle\pi\,\delta^{(3)}(r)\rangle_{\gamma}}\,A_{\gamma}\bigg]^\mathrm{theo}\,. \label{eq:40}
\end{align}
The first term is the atomic hyperfine splitting scaled by the contact interaction. Namely, 
the denominator in the first term of Eq.~(\ref{eq:40}) is
\begin{align}
\langle\pi\,\delta^{(3)}(r)\rangle_{\gamma} =&\ \Big(\frac{m}{M_\gamma}+1\Big)^{-3}\,, \label{eq:41}
\end{align}
and the numerator $\langle\pi\,\delta^{(3)}(r_{e\gamma})\rangle_{vL}$ 
is the expectation value with the nonadiabatic (full three-body) wave function.
The second term, the difference between HD$^+$ and H (D), is calculated by an expansion in $\alpha$.
The two leading terms in $\alpha$ vanish in this difference, because they are proportional to 
$\langle\pi\,\delta^{(3)}(r_{e\gamma})\rangle$. The nonvanishing terms are the state-dependent contributions $A^{(6)}_{e\gamma}$ and  $A^{(7)}_{e\gamma}$,
and higher-order terms in $\alpha$ are neglected. The $A^{(6)}_{e\gamma}$ and  $A^{(7)}_{e\gamma}$ terms are calculated
in the BO approximation, so that terms quadratic in the electron-to-nucleus mass ratio are also neglected. 
The $A_{e\gamma}$ couplings for a given rovibrational state in the BO approximation are obtained from Eq.~(\ref{eq:40}) as
\begin{align}
A_{e\gamma}(v,L) =&\
\frac{\langle\pi\,\delta^{(3)}(r_{e\gamma})\rangle_{vL}}{\langle\pi\,\delta^{(3)}(r)\rangle_{\gamma}}\,A_{\gamma}^\mathrm{exp}
\nonumber \\ &\ +
\big[ \Delta^{(6)}_{vL} + \delta^{(6)}_{vL}\big]\,(Z\,\alpha)^6\,\frac{m^2}{M_\gamma}\,g_\gamma
\nonumber \\ &\
+ \big[\Delta^{(7)}_{vL} + \delta^{(7)}_{vL}\big]\,\frac{\alpha}{\pi}\,(Z\,\alpha)^6\,\frac{m^2}{M_\gamma}\,g_\gamma
\nonumber \\ \equiv&\
\bar A_{e\gamma}(v,L) + A^{(6)}_{e\gamma}(v,L) + A^{(7)}_{e\gamma}(v,L)\,, \label{eq:42}
\end{align}
where we keep $Z$ arbitrary only to identify the order of the correction, and
\begin{align}
\Delta^{(6)}_{vL} =&\ \big\langle v,L\bigl|{\cal A}^{(6)}_{e\gamma}(R)
-\langle\pi\,\delta^{(3)}(r_{\gamma})\rangle\,{\cal A}^{(6)}_H\big|v,L\big\rangle\,, \label{eq:43}
\\
\Delta^{(7)}_{vL} =&\ \big\langle v,L\bigl|{\cal A}^{(7)}_{e\gamma}(R)
-\langle\pi\,\delta^{(3)}(r_{\gamma})\rangle\,{\cal A}^{(7)}_H\big|v,L\big\rangle\,, \label{eq:44}
\\
\delta^{(6)}_{vL} =&\  2\,\bigg\langle v,L\biggl|{\cal E}^{(4)}(R)\,\frac{1}{({\cal E}_{vL} -{\cal H}_\mathrm{BO})'}\,{\cal A}^{(4)}_{e\gamma}(R) \bigg|v,L\bigg\rangle\,, \label{eq:45}
\\
\delta^{(7)}_{vL} =&\
2\,\bigg\langle v,L\biggl|{\cal E}^{(5)}(R)\,\frac{1}{({\cal E}_{vL} -{\cal H}_\mathrm{BO})'}\,{\cal A}^{(4)}_{e\gamma}(R) \bigg|v,L\bigg\rangle
\nonumber \\ &\
+\frac{1}{2}\, \delta^{(6)}_{vL}\,, \label{eq:46}
\end{align}
where the last term comes from the electron magnetic-moment anomaly, $\langle\pi\,\delta^{(3)}(r_{\gamma})\rangle$ is the electronic expectation value at fixed $R$,
and ${\cal A}^{(6)}_H$ and ${\cal A}^{(7)}_H$ are the corresponding coefficients for the hydrogen atom in the nonrecoil limit.

\section{Hyperfine splitting in hydrogen-like atoms}\label{sec:hydrogen}
Let us start the derivation of ${\cal A}^{(6)}_{e\gamma}$ and ${\cal A}^{(7)}_{e\gamma}$ 
by summarizing the existing theory of the magnetic-dipole hfs of the $1S$ state
of hydrogen-like atoms. To leading order in the fine-structure constant $\alpha$, it is given by
\begin{align}
A^{(4)} = E_F =
\frac{4}{3}(Z\alpha)^4\frac{m_\mathrm{r}^3}{m\,M}\,g\,, \label{eq:47}
\end{align}
where $m_\mathrm{r} = mM/(m+M)$ is the reduced mass,
$m$ and $M$ are the masses of the electron and the nucleus,
respectively, and $Z$ is the nuclear charge number.
The leading-order hfs contribution in Eq.~(\ref{eq:47}) is also known as the Fermi energy.
The complete hyperfine structure of $S$ states in hydrogenic systems
is represented as an expansion in terms of $\alpha$,
\begin{align}
A =&\ E_F\,(1 + \delta)\,, \label{eq:48}
\end{align}
where $\delta$ incorporates various corrections. Here we include only the relativistic correction $\delta^{(2)}_H$ and the one-loop radiative correction $\delta^{(3)}_H$,
which exhibit a significant state dependence and thus contribute to $\Delta$ in Eq.~(\ref{eq:42}),
\begin{align}
\delta  =&\
\kappa+ \delta^{(2)}_H+ \delta^{(3)}_H\,. \label{eq:49}
\end{align}
Here, $\kappa $ is the magnetic-moment anomaly of the free electron, $\kappa = \alpha/(2\pi) + O(\alpha^2)$, and \cite{eides:01}
\begin{align}
\delta^{(2)}_H&\  =\frac{3}{2}\,(Z\,\alpha)^2\,, \label{eq:50}                \\
\delta^{(3)}_H&\  = \frac{\alpha\,(Z\,\alpha)^2}{\pi}\!
\bigg[\!-\frac{8}{3}\ln(Z\,\alpha)\Bigl(\ln(Z\,\alpha)-\ln 4+\frac{281}{480}\Bigr)
\nonumber \\ &\
+ 17.122\,338\,751\,3-\frac{8}{15}\,\ln 2+\frac{34}{225}\biggr]\,, \label{eq:51}
\end{align}
where the first term in the second line comes from the self-energy \cite{krp:96, patkos:23} and the second one from the vacuum polarization.
Below we derive the corresponding corrections $A^{(6)}_{e\gamma}$ and $A^{(7)}_{e\gamma}$ for HD$^+$
and verify that their large-$R$ limits coincide with half of the hydrogen values.

\section{Relativistic correction of order $\alpha^6$}\label{sec:A6}
The relativistic $\alpha^2\,E_F$ contribution to the scalar hyperfine couplings is 
\begin{align}
{E}^{(6)}_{\rm hfs} =&\
\langle H^{(6)}_{\rm hfs}\rangle
+2\,\langle H^{(4)}\,\frac{1}{({E}-H)'}\,H^{(4)}_{\rm hfs}\rangle\,. \label{eq:52}
\end{align}
The hyperfine Hamiltonian is
\begin{align}
H^{(4+)}_{\rm hfs} =&\ H^{(4+)}_{{\rm hfs},A} + H^{(4)}_{{\rm hfs},B}\,, \label{eq:53}                 \\
H^{(4+)}_{{\rm hfs},A} =&\ \frac{2\,Z\,\alpha}{3\,m}\,(1+\kappa)\sum_\gamma
\frac{g_\gamma}{M_\gamma}\,\vec{I}_\gamma\cdot\vec{\sigma}\,\pi\,\delta^{(3)}(r_\gamma)\,, \label{eq:54}                 \\
H^{(4)}_{{\rm hfs},B} =&\ \frac{Z\,\alpha}{2\,m}\,\sum_\gamma
\frac{g_\gamma}{M_\gamma}\,\vec{I}_\gamma\cdot\frac{\vec{r}_\gamma}{r_\gamma^3}\times\vec{p}\,, \label{eq:55}
\end{align}
where $\vec I_\gamma\equiv \vec s_\gamma$, $\kappa$ is the electron magnetic-moment anomaly,
and $H^{(4+)}$ is the Breit Hamiltonian in the nonrecoil limit,
\begin{align}
H^{(4+)} =&\  H^{(4)}_A+H^{(4+)}_B\,, \label{eq:56}                 \\
H^{(4+)}_B =&\ \sum_\gamma\frac{Z\,\alpha}{4\,m^2}\,
\frac{\vec{r}_\gamma}{r_\gamma^3}\times\vec{p}\cdot\vec{\sigma}\,(1+2\,\kappa)\,, \label{eq:57}
\end{align}
and $H^{(4)}_A$ is the operator ${\cal H}^{(4)}_A$ of Eq.~(\ref{eq:23}) written in natural units.

The operator $H_{\rm hfs}^{(6)}$ in Eq.~(\ref{eq:52}) includes nuclear-spin-dependent operators
that contribute at order $m\,\alpha^6$. 
In hydrogenic systems it leads to $\delta^{(2)}_H$ in Eq.~(\ref{eq:50}). 
Its derivation for HD$^+$ proceeds as follows. The vector potential and the magnetic field due to the magnetic moment $\vec\mu_\gamma$ of the nucleus $\gamma$ are
\begin{align}
e\,\vec A_\gamma(\vec r) = &\ \frac{e}{4\pi}\,\vec{\mu}_\gamma\times\frac{\vec r_\gamma}{r_\gamma^3}
= -Z\alpha\,\frac{g_\gamma}{2\,M_\gamma} \vec I_\gamma\times \frac{\vec r_\gamma}{r_\gamma^3}\,, \label{eq:58}                \\
e\,B_\gamma^i(\vec r) = &\ e\,(\nabla\times\vec A_\gamma)^i = -Z \alpha\, \frac{g_\gamma}{2\,M_\gamma}\frac{8\pi}{3}\delta^{(3)}(r_\gamma)\,I_\gamma^i
\nonumber \\ &\
+ Z\alpha \,\frac{g_\gamma}{2\,M_\gamma}\frac{1}{r_\gamma^3} \bigg(\delta^{ij} - 3\frac{r_\gamma^i r_\gamma^j}{r_\gamma^2}\bigg) I_\gamma^j\,. \label{eq:59}
\end{align}
Consider the part of the Foldy--Wouthuysen (FW) Hamiltonian \cite{lwqed} that includes the coupling of the electron spin to the magnetic field,
\begin{align}
\delta H_{\rm FW} =&\
\frac{\vec\pi^2}{2\,m}-\frac{e}{2\,m}\,\vec\sigma\cdot\vec B
\nonumber \\ &\
+\sum_\beta\frac{Z\,\alpha}{4\,m^2}\,\vec\sigma\cdot\frac{\vec r_\beta}{r_\beta^3}\times\vec\pi
+\frac{e}{8\,m^3}\,\bigl\{\vec\sigma\cdot\vec B\,,\,\vec \pi^2\big\}\,, \label{eq:60}
\end{align}
where $\vec\pi =\vec p - e\,\vec A$, $\vec A = \sum_\gamma \vec A_\gamma$, and $\vec B = \sum_\gamma \vec B_\gamma$. 
The leading interaction $H^{(4)}_{\rm hfs}$
between the nuclear spins $\vec I_\gamma$ and the electron spin $\vec s_e = \vec \sigma/2$
is obtained from the nonrelativistic terms [the first line of Eq.~(\ref{eq:60})],
\begin{equation}
H^{(4)}_{\rm hfs} = -\sum_\gamma \biggl[\frac{e}{m}\,\vec p\cdot\vec A_\gamma(\vec r)
+\frac{e}{2\,m}\,\vec\sigma\cdot\vec B_\gamma(\vec r)\biggr]\,, \label{eq:61}
\end{equation}
with the vector potential and the magnetic field of the nuclei given by Eqs.~(\ref{eq:58}) and (\ref{eq:59}).
The relativistic correction $H^{(6)}_{\rm hfs}$ is similarly obtained from $\delta H_{\rm FW}$ [the second line of Eq.~(\ref{eq:60})],
\begin{align}
H^{(6)}_{\rm hfs} =&\
\sum_\gamma\biggl[\frac{(Z\,\alpha)^2}{12\,m^2}\,\frac{\vec r_\gamma}{r_\gamma^3}\cdot\biggl(\frac{\vec r_A}{r_A^3} + \frac{\vec r_B}{r_B^3}\biggr)
\nonumber \\ &\
-\frac{Z\,\alpha}{6\,m^3}\,
\bigl\{p^2\,,\,\pi\,\delta^{(3)}(r_\gamma)\bigr\}\biggr]
\frac{g_\gamma}{M_\gamma}\,\vec\sigma\cdot\vec I_\gamma\,, \label{eq:62}
\end{align}
in agreement with the earlier, but much lengthier, derivation in Ref.~\cite{korobov:20}. The total result is
\begin{align}
A^{(6)}_{e\gamma}(R) =&\ {\cal A}^{(6)}_{e\gamma}(R)\, (Z\,\alpha)^6\,\frac{m^2}{M_\gamma}\,g_\gamma\,, \label{eq:63}                 \\
{\cal A}^{(6)}_{e\gamma}(R) =&\ {\cal A}^{(6)}_{e\gamma 1} + {\cal A}^{(6)}_{e\gamma 2} + {\cal A}^{(6)}_{e\gamma 3}\,, \label{eq:64}                 \\
{\cal A}^{(6)}_{e\gamma 1} =&\
\biggl\langle\frac{1}{6}\,\frac{\vec r_\gamma}{r_\gamma^3}\cdot\biggl(\frac{\vec r_A}{r_A^3} + \frac{\vec r_B}{r_B^3}\biggr)
-\frac{1}{3}\,\bigl\{p^2\,,\,\pi\,\delta^{(3)}(r_\gamma)\bigr\}\biggr\rangle\,, \label{eq:65}                 \\
{\cal A}^{(6)}_{e\gamma 2}  =&\
\frac{2}{3}\,\biggl\langle 4\,\pi\,\delta^{(3)}(r_\gamma)\,\frac{1}{({\cal E}-{\cal H})'}
\nonumber \\ &\times
\bigg(-\frac{p^4}{8}+\frac{\pi}{2}\,[\delta^{(3)}(r_A) + \delta^{(3)}(r_B)]\bigg)\biggr\rangle\,, \label{eq:66}                \\
{\cal A}^{(6)}_{e\gamma 3}  =&\
\frac{1}{6}\,\biggl\langle \,\frac{\vec{r}_\gamma}{r_\gamma^3}\times\vec{p}\,\frac{1}{{\cal E}-{\cal H}}\,
\bigg(\frac{\vec{r}_A}{r_A^3} + \frac{\vec{r}_B}{r_B^3}\bigg)\times\vec{p}\biggr\rangle\,. \label{eq:67}
\end{align}
However, the above matrix elements are singular. We obtain a finite expression
by transforming the operators in the second-order matrix elements according to
\begin{align}
{\cal H}^{(4)}_A \equiv&\ {\cal H}'^{(4)}_A + \frac{1}{4}\,\biggl\{\frac{1}{r_A} + \frac{1}{r_B}\,,\,{\cal E}-{\cal H}\biggr\}\,, \label{eq:68}                 \\
4\,\pi\,\delta^{(3)}(r_\gamma)\equiv&\ 4\,\pi\,[\delta^{(3)}(r_\gamma)]' -
\biggl\{\frac{2}{r_\gamma}\,,\,{\cal E}-{\cal H}\biggr\}\,, \label{eq:69}
\end{align}
where the regularized operators are such that
\begin{align}
4\,\pi\,[\delta^{(3)}(r_\gamma)]' |\phi\rangle =&\ -2\,\frac{\vec r_\gamma}{r_\gamma^3}\cdot\vec\nabla|\phi\rangle\,, \label{eq:70}                 \\
{\cal H}'^{(4)}_A|\phi\rangle =&\  \bigg[-\frac{1}{2}\,\bigg({\cal E} +\frac{1}{r_A} + \frac{1}{r_B}\bigg)^2
\nonumber \\ &\
-\frac{1}{4}\,\bigg(\frac{\vec r_A}{r_A^3} + \frac{\vec r_B}{r_B^3}\bigg)\cdot\vec\nabla\bigg]|\phi\rangle\,. \label{eq:71}
\end{align}
All singular terms are moved to the first-order matrix elements,
which, when combined, form a well-defined and finite expression,
\begin{widetext}
\begin{align}
{\cal A}^{(6)}_{e\gamma}=&\
\frac{1}{3}\,\biggl\langle \vec p\,\frac{1}{r_\gamma}\,\bigg({\cal E}+ \frac{1}{r_A} + \frac{1}{r_B}\bigg)\vec p\,\biggr\rangle
-\frac{2}{3}\,\langle \pi\,\delta^{(3)}(r_\gamma)\rangle\,\biggl\langle {\cal E} + \frac{1}{r_A} + \frac{1}{r_B}\biggr\rangle
+\frac{4}{3}\,\biggl\langle \frac{1}{r_\gamma}\bigg\rangle\, {\cal E}^{(4)}
\nonumber \\ &\
+\frac{2}{3}\,\biggl\langle 4\,\pi\,[\delta^{(3)}(r_\gamma)]' \,\frac{1}{({\cal E}-{\cal H})'}\,{\cal H}'^{(4)}_A\biggr\rangle
+\frac{1}{6}\,\biggl\langle \,\frac{\vec{r}_\gamma}{r_\gamma^3}\times\vec{p}\,\frac{1}{{\cal E}-{\cal H}}\,
\bigg(\frac{\vec{r}_A}{r_A^3} + \frac{\vec{r}_B}{r_B^3}\bigg)\times\vec{p}\biggr\rangle\,. \label{eq:72}
\end{align}
\end{widetext}
Finally, the large-$R$ limit of ${\cal A}^{(6)}_{e\gamma}(R)$ is
\begin{align}
\lim_{R\rightarrow\infty}{\cal A}^{(6)}_{e\gamma}(R) = \frac{1}{2}\,{\cal A}^{(6)}_H = 1\,, \label{eq:73}
\end{align}
which is verified numerically in Table~\ref{tab:VIII}. In the previous calculation of Ref.~\cite{korobov:20}
the last term in Eq.~(\ref{eq:72}) was omitted.

\section{Radiative correction of order $\alpha^7$}\label{sec:A7}
The one-loop self-energy and vacuum-polarization corrections to the hfs of order $\alpha^7$
can be represented as
\begin{align}
E^{(7)}_\mathrm{hfs} =&\   E^{(7)}_L + E^{(7)}_\mathrm{sec} + \langle H^{(7)}_{\rm hfs}\rangle
\nonumber \\ =&\ \sum_\gamma A^{(7)}_{e\gamma}\,\vec I_\gamma\cdot\vec s_e\,, \label{eq:74}
\end{align}
where $E^{(7)}_L$ is the low-energy contribution with $k\sim m\,\alpha^2$, $E^{(7)}_\mathrm{sec}$ is the second-order contribution
coming from the lower-order effective Hamiltonians $H^{(4)}$ and $H^{(5)}$, and $H^{(7)}_{\rm hfs}$ is a higher-order effective QED Hamiltonian.
The derivation of all three parts is presented in Appendix~\ref{app:A7}.
They contain the photon-momentum cutoff $\lambda$, which cancels out in the sum in Eq.~(\ref{eq:74}). Whenever convenient, we use $A^{(7)}$ or ${\cal A}^{(7)}$ instead of $E^{(7)}$.
After the elimination of $\lambda$ and of the divergences, and after matching with the hydrogenic result, 
${\cal A}^{(7)}_{e\gamma}$ is the sum of three finite terms,
\begin{align}
{\cal A}^{(7)}_{e\gamma} =&\ {\cal A}^{(7)}_{L'} + {\cal A}^{(7)}_\mathrm{sec} + {\cal A}^{(7)}_\mathrm{fo}\,, \label{eq:75}
\end{align}
namely the low-energy contribution, the second-order contribution, and the first-order contribution.
The first-order contribution includes matrix elements that require a specific definition,
\begin{align}
\biggl\lbr\frac{1}{\rg^3}\biggr\rbr =&\ \lim_{\epsilon\to0}\biggl[\int_{\rg>\epsilon} d^3r\,\frac{\phi^2}{\rg^3}
\nonumber \\ &\ + 4\,\lbr\pi\,\dg\rbr\,(\gamma_E+\ln\epsilon)\biggr]\,, \label{eq:76}                \\
\biggl\lbr\frac{1}{\rg^4}\biggr\rbr =&\ \lim_{\epsilon\to0}\biggl[\int_{\rg>\epsilon} d^3r\,\frac{\phi^2}{\rg^4}
\nonumber \\ &\ - \frac{4\,\lbr\pi\,\dg\rbr}{\epsilon} - 8\,\lbr\pi\dg\rbr\,(\gamma_E+\ln\epsilon)\biggr]\,. \label{eq:77}
\end{align}
Their large-$R$ limits are
\begin{align}
\lim_{R\rightarrow\infty}\biggl\lbr\frac{1}{\rg^3}\biggr\rbr =&\  \frac{1}{2}\,  \biggl\lbr\frac{1}{r^3}\biggr\rbr_H = -2\,\ln 2\,, \label{eq:78}           \\
\lim_{R\rightarrow\infty}\biggl\lbr\frac{1}{\rg^4}\biggr\rbr =&\   \frac{1}{2}\,  \biggl\lbr\frac{1}{r^4}\biggr\rbr_H = 4\,(\ln 2-1)\,. \label{eq:79}
\end{align}
Moreover, for $\phi(\vec r) = \phi(r_A,r_B)$ we define $\langle \vec p\;\delta^{(3)}(r_A)\,\vec p\,\rangle$ as
\begin{align}
 \langle \vec p\;\delta^{(3)}(r_A)\,\vec p\,\rangle =&\ \bigg(\frac{\partial\phi}{\partial r_B}\biggr)^2\bigg|_{r_A=0, r_B=R}\,, \label{eq:80}
\end{align}
which vanishes in the hydrogenic and large-$R$ limits.
The first-order contribution ${\cal A}^{(7)}_\mathrm{fo}$, after matching with the hydrogenic value, is
\begin{widetext}
\begin{align}
{\cal A}^{(7)}_\mathrm{fo} =&\
\bigg\langle\frac{\vec r_\gamma}{r_\gamma^3}\cdot\biggl(\frac{\vec r_A}{r_A^3} + \frac{\vec r_B}{r_B^3}\biggr)\bigg\rangle\,
\bigg(\frac{4}{9}\,\tau + \frac{7}{24}\bigg)
+  \biggl\langle \frac{1}{r_\gamma}\,\bigg({\cal E}^{(2)}+\frac{1}{r_A} + \frac{1}{r_B}-\frac{1}{R}\bigg)^2\biggr\rangle\,\frac{1}{3}
\nonumber \\ &\
+\bigg\langle \frac{1}{r_\gamma}\bigg\rangle\,{\cal E}^{(4)}\,\frac{2}{3}
+\bigg\langle \frac{1}{r_A}+\frac{1}{r_B}\bigg\rangle\,\langle\pi\,\delta^{(3)}(r_\gamma)\rangle\,\bigg(\frac{32}{9}\,\tau -\frac{1}{3}\bigg)
- {\cal E}^{(2)}\,\langle \pi\,\delta^{(3)}(r_\gamma)\rangle\,\bigg(\frac{16}{9}\,\tau + \frac{5}{18}\bigg)
\nonumber \\ &\
+ \langle \vec p\,\pi\,\delta^{(3)}(r_\gamma)\,\vec p\,\rangle\,\bigg(\frac{8}{9}\,\tau -\frac{1}{36}\bigg)
- \frac{1}{R}\,\langle\pi\delta^{(3)}(r_\gamma)\rangle\,\frac{32}{9}\,\tau
\nonumber \\ &\
+ \langle\pi\delta^{(3)}(r_\gamma)\rangle\bigg[
-\frac{5351}{2700} - \frac{22\pi^2}{27} - \frac{5}{3}\zeta(3) + \frac{448}{27}\ln 2
+ \frac{8}{9}\ln^2 2  - \frac{2441}{135}\ln\alpha - \frac{32}{9}\ln^2\alpha + \frac{128}{9}\ln 2\ln\alpha\bigg]\,, \label{eq:81}
\end{align}
where
\begin{align}
\tau =&\ 5/6 -1/5 +\ln(\alpha^{-2}/2)\,. \label{eq:82}
\end{align}
The second-order contribution ${\cal A}^{(7)}_\mathrm{sec}$ is
\begin{align}
{\cal A}^{(7)}_\mathrm{sec} =&\
\frac{2}{9}\,\tau\,\bigg\langle 4\,\pi\,[\delta^{(3)}(r_\gamma)]'\,\frac{1}{({\cal E}-{\cal H})'}\,4\,\pi\,[\delta^{(3)}(r_A) + \delta^{(3)}(r_B)]'\bigg\rangle
+ \frac{1}{3}\,\biggl\langle 4\,\pi\,[\delta^{(3)}(r_\gamma)]'\,\frac{1}{({\cal E}-{\cal H})'}\,{\cal H}'^{(4)}_A\biggr\rangle
\nonumber \\ &\
+ \frac{1}{6}\,\biggl\langle \,\frac{\vec{r}_\gamma}{r_\gamma^3}\times\vec{p}\,\frac{1}{{\cal E}-{\cal H}}\,
\bigg(\frac{\vec{r}_A}{r_A^3} + \frac{\vec{r}_B}{r_B^3}\bigg)\times\vec{p}\biggr\rangle\,. \label{eq:83}
\end{align}
The low-energy contribution ${\cal A}^{(7)}_{L'}$ of Eq.~(\ref{eq:A10}) is
\begin{align}
{\cal A}^{(7)}_{L'} =&\
\frac{2}{3}\,\bigg[ \int_0^1 dt\,\frac{f_\gamma(t) - a_\gamma-\sqrt{2}\,b_\gamma\,t+2\,c_\gamma\,\ln(2\,t^2)\,t^2-2\,d_\gamma\,t^2}{t^3}
-\sqrt{2}\, b_\gamma-\frac{c_\gamma}{2}\,\ln^2 2+d_\gamma\,\ln 2 \bigg]\,, \label{eq:84}
\end{align}
where the function $f_\gamma(t)$ is defined by Eq.~(\ref{eq:A3}) with $t=1/\sqrt{1+2\,k}$, and the coefficients $a_\gamma$, $b_\gamma$, $c_\gamma$, and $d_\gamma$ of its large-$k$ expansion are given in Eqs.~(\ref{eq:A5})--(\ref{eq:A8}).
Their large-$R$ limits are equal to half of the hydrogenic values \cite{krp:96, patkos:23},
\begin{align}
\lim_{R\rightarrow\infty} {\cal A}^{(7)}_{L'}(R) =&\  \frac{2}{3}\bigg[ n_1 + n_2 + \frac{781}{18} +  \frac{4}{3}\,\pi^2 - \frac{166}{3}\,\ln 2 - \frac{2}{3}\,\ln^2 2\bigg]\,, \label{eq:85}           \\
\lim_{R\rightarrow\infty} {\cal A}^{(7)}_\mathrm{sec}(R) =&\ \frac{2}{3}\bigg[-\frac{125}{36} + \frac{20}{3}\,\ln 2 + \frac{40}{3}\,\ln\alpha\bigg]\,, \label{eq:86}           \\
\lim_{R\rightarrow\infty} {\cal A}^{(7)}_\mathrm{fo}(R) =&\ \frac{2}{3}\bigg[
-\frac{173}{50} - \frac{11}{18}\,\pi^2 + \frac{853}{60}\,\ln 2 - 2\,\ln^2 2 - \frac{2681}{180}\,\ln\alpha +
   \frac{16}{3}\,\ln 2\,\ln\alpha - \frac{8}{3}\,\ln^2\alpha - \frac{5}{4}\,\zeta(3)\bigg]\,, \label{eq:87}           \\
\lim_{R\rightarrow\infty} {\cal A}^{(7)}_{e\gamma}(R) =&\  \frac{2}{3}\,
\bigg[\!-\frac{8}{3}\ln\alpha\Bigl(\ln\alpha-2\,\ln 2+\frac{281}{480}\Bigr)+ 17.122\,338\,751\,3-\frac{8}{15}\,\ln 2+\frac{34}{225}\biggr]\,, \label{eq:88}
\end{align} 
\end{widetext}
where $n_1$ and $n_2$ are taken from Refs.~\cite{krp:96,patkos:23},
\begin{align}
n_1 =&\ -0.085\,740\,323\,701\,4\,, \label{eq:89}            \\
n_2 =&\  \hspace{12pt} 0.067\,496\,936\,500\,3\,. \label{eq:90}
\end{align}
The above asymptotics are verified numerically in Table~\ref{tab:VIII} in Appendix~\ref{app:num}.

\section{Numerical calculations}\label{sec:num}
\begin{figure}[tb]
\includegraphics[width=\columnwidth]{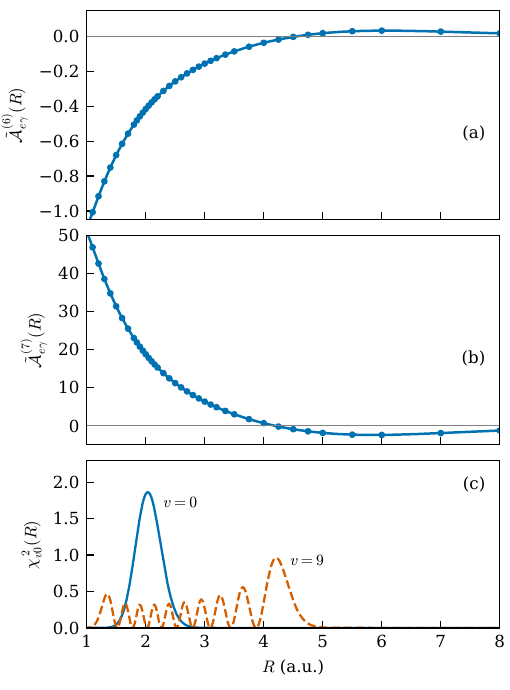}
\caption{\label{fig:potentials} (a), (b) The functions $\tilde{\cal A}^{(6)}_{e\gamma}(R)$ and $\tilde{\cal A}^{(7)}_{e\gamma}(R)$ of Eqs.~(\ref{eq:91}) and (\ref{eq:92}),
i.e., the relativistic and radiative corrections to the scalar coupling with the hydrogenic part subtracted. 
The points are the calculated values from Table~\ref{tab:VIII}, and the lines are the interpolation used for the rovibrational averaging. 
Both functions vanish for $R\to\infty$. 
(c) The probability densities $\chi^2_{vL}(R)$ of the internuclear distance for $L=0$ and $v=0$ (solid line) and $v=9$ (dashed line).}
\end{figure}

The numerical calculations of $A_{e\gamma}$ are performed according to Eq.~(\ref{eq:42}).
The first term, $\bar A_{ep}$ ($\bar A_{ed}$), is the rescaled atomic hydrogen (deuterium) hyperfine splitting. It involves the electron--nucleus Dirac $\delta$ function,
whose expectation value with the nonadiabatic wave function is recalculated here using the CODATA 2022 nuclear masses; see Ref.~\cite{pantak:26} for details and Refs.~\cite{Korobov:19,zhong:12} for the previous calculations. 
The second and third terms in Eq.~(\ref{eq:42}) are calculated in the BO approximation.  
Namely, ${\cal E}^{(2)}$, ${\cal E}^{(4)}$, ${\cal E}^{(5)}$, ${\cal A}^{(4)}_{e\gamma}$, ${\cal A}^{(6)}_{e\gamma}$, and ${\cal A}^{(7)}_{e\gamma}$
are calculated on a selected grid of the proton--deuteron distance $R$, 
and, in addition, ${\cal E}^{(2)}$, ${\cal E}^{(4)}$, and ${\cal A}^{(4)}_{e\gamma}$ are calculated on the fly for the averaging with the radial function $\chi_{vL}(R)$.
Numerically, the most demanding part is the calculation of the Bethe logarithm \cite{korobov:04,korobov:06,korobov:13} in ${\cal E}^{(5)}$ and ${\cal A}^{(7)}_{e\gamma}$.
We found it numerically efficient to replace the integral over the photon momentum $k$ by an integral over $t =(1+2\,k)^{-1/2}$;
see Appendix~\ref{app:num} for details of the numerical calculations.
We have checked analytically and numerically that the large-$R$ asymptotics of ${\cal E}^{(5)}$, ${\cal A}^{(6)}_{e\gamma}$, and ${\cal A}^{(7)}_{e\gamma}$ 
agree with the hydrogenic values within the quoted numerical precision. 
Next, we consider the differences
\begin{align}
\tilde {\cal A}^{(6)}_{e\gamma}(R) =&\ {\cal A}^{(6)}_{e\gamma}(R) -\langle\pi\,\delta^{(3)}(r_{\gamma})\rangle\,{\cal A}^{(6)}_H\,, \label{eq:91}
\\
\tilde {\cal A}^{(7)}_{e\gamma}(R) =&\ {\cal A}^{(7)}_{e\gamma}(R) -\langle\pi\,\delta^{(3)}(r_{\gamma})\rangle\,{\cal A}^{(7)}_H\,, \label{eq:92}
\end{align}
in which many terms cancel out (see Fig.~\ref{fig:potentials}), and calculate their expectation values according to
\begin{align}
\Delta^{(6)}_{vL} =&\ \big\langle v,L\bigl|\tilde {\cal A}^{(6)}_{e\gamma}(R) \big|v,L\big\rangle\,, \label{eq:93}          \\
\Delta^{(7)}_{vL} =&\ \big\langle v,L\bigl|\tilde{\cal A}^{(7)}_{e\gamma}(R)  \big|v,L\big\rangle\,, \label{eq:94}
\end{align}
with radial wave functions obtained using the discrete variable representation (DVR) \cite{colbert:92}.
Similarly, $\delta^{(6)}_{vL}$ and $\delta^{(7)}_{vL}$ are obtained using a DVR representation
of the Hamiltonian ${\cal H}_\mathrm{BO}$, with ${\cal E}^{(4)}(R)$ calculated on demand and ${\cal E}^{(5)}(R)$ interpolated from the grid values. 
The numerical results for $A_{ep}$ and $A_{ed}$ are presented in Table~\ref{tab:I}, while the technical details 
of our numerical calculations and intermediate numerical results are presented in Appendix~\ref{app:num}. 
The uncertainties in parentheses account for the unknown higher-order terms of Eq.~(\ref{eq:42}) and are estimated 
as $2\,g_\gamma\,m/M_\gamma\,|A^{(6)}_{e\gamma}|+\alpha\,|A^{(7)}_{e\gamma}|$, 
with $A^{(6)}_{e\gamma}$ and $A^{(7)}_{e\gamma}$ from Tables~\ref{tab:X} and \ref{tab:XI};
they are much larger than our numerical uncertainties.
The calculation of these higher-order terms is within reach and would allow one, in the near future,
to reduce the uncertainty of the scalar hyperfine couplings in the HD$^+$ and H$_2^+$ molecular ions below 10 Hz.

\section{Comparison with previous calculations}\label{sec:karr}
In Table~\ref{tab:VII} we compare the complete $(Z\,\alpha)^2\,E_F$ correction
$\big[\langle {\cal A}^{(6)}_{e\gamma}\rangle_{vL}+\delta^{(6)}_{vL}\big] (Z\,\alpha)^6\,m^2\,g_\gamma/M_\gamma$ 
for the relevant rovibrational states $(v,L)=(0,0)$, $(0,3)$, and $(9,3)$, with that of Karr \emph{et al.}~\cite{karr:20}.
They included the rovibrational contribution $\delta^{(6)}_{vL}$ of Eq.~(\ref{eq:45}),
but neglected the second-order spin--orbit term ${\cal A}^{(6)}_{e\gamma 3}$ of Eq.~(\ref{eq:67}).
As shown in Table~\ref{tab:VII}, omitting this term brings our results into agreement with
Ref.~\cite{karr:20} to about $0.05$~kHz for $A_{ep}$ and $0.007$~kHz for $A_{ed}$,
comparable to their numerical precision and also to the uncertainty due to the BO approximation in our calculations.
\begin{table}[htb]
\caption{\label{tab:VII} Relativistic $(Z\,\alpha)^2E_F$ contributions to the HD$^+$ scalar couplings in kHz.
The results of Ref.~\cite{karr:20} are compared with our calculations without (w/o) and with the second-order
spin--orbit contribution ${\cal A}^{(6)}_{e\gamma 3}$ of Eq.~(\ref{eq:67}); the latter is our complete result without subtraction
of the hydrogenic term ${\cal A}^{(6)}_H$.}
\begin{ruledtabular}
\begin{tabular}{cd{2.1}d{2.3}d{2.3}d{2.2}d{2.3}d{2.3}}
 & \multicolumn{3}{c}{$A^{(6)}_{ep}$} & \multicolumn{3}{c}{$A^{(6)}_{ed}$} \\
$(v,L)$ & \centt{Ref.~\cite{karr:20}} & \centt{w/o} & \centt{with} & \centt{Ref.~\cite{karr:20}} & \centt{w/o} & \centt{with} \\ \hline
$(0,0)$ & 66.9 & 66.950 & 66.300 & 10.27 & 10.277 & 10.177 \\
$(0,3)$ & 66.5 & 66.466 & 65.825 & 10.20 & 10.203 & 10.105 \\
$(9,3)$ & 57.2 & 57.243 & 56.897 & 8.79 & 8.787 & 8.734 \\
\end{tabular}
\end{ruledtabular}
\end{table}

The total difference from the previous results of Ref.~\cite{karr:20} originates from the complete calculation of the relativistic 
and radiative corrections, in particular from the inclusion of the second-order spin–orbit contribution and 
the full state-dependent radiative correction $\tilde {\cal A}^{(7)}_{e\gamma}$.

\section{Summary and outlook}\label{sec:summary}
We have calculated the scalar spin interactions $A_{ep}\,\vec s_e\cdot\vec s_p$ and $A_{ed}\,\vec s_e\cdot\vec s_d$
in the HD$^+$ molecular ion using the NRQED approach and the experimental values of the hydrogen and deuterium hyperfine splittings, 
according to Eq.~(\ref{eq:40}). We rescaled the measured atomic hfs by the ratio of the electron densities at the nucleus and calculated the state-dependent remainder
in the BO approximation. More precisely,
we have obtained closed formulas for $A^{(6)}_{e\gamma}$ and $A^{(7)}_{e\gamma}$
and calculated them numerically in the clamped-nuclei approximation on a large grid of internuclear distances. 
Next, we used the DVR method to calculate $A_{e\gamma}$ for the rovibrational states with
$0\leq v \leq 9$ and $0\leq L \leq 3$. 

Our numerical results in Table~\ref{tab:I} are 18 times more accurate than the previous results 
of Karr {\em et al.} \cite{karr:20} and differ by up to about three of their standard deviations.
Moreover, our calculations are in agreement with the measurements of the hyperfine splittings in the $(v,L) = (0,0)$ \cite{konig:26} and
$(v,L) = (0,3)$ \cite{kliukin:26} states,
but disagree with the measurement \cite{patra:20} for the $(v,L) = (0,3)\rightarrow (9,3)$ transition.
Since the hyperfine structure of the $(0,3)$ state is now confirmed experimentally, the remaining discrepancy of $7.4(9)$~kHz may originate 
from an unaccounted contribution to the hyperfine structure of the $(9,3)$ state or from an experimental systematic effect in the overtone measurement. 
Its explanation requires an independent measurement of a hyperfine interval within this state 
and a verification of the spin–orbit and tensor couplings, which dominate the present theoretical uncertainty.

The new results for $A_{e\gamma}$ provide improved input for extracting spin-averaged frequencies and determining mass
ratios from measurements of transition frequencies between rovibrational states of HD$^+$ \cite{patra:20,kortunov:21,alighanbari:23}.
Moreover, Table~\ref{tab:VIII} provides the data needed to calculate the scalar hyperfine couplings in all other isotopologues;
these have to be supplemented by the nuclear $g$ factors and masses and by the finite-mass (nonadiabatic) contact densities.
 
The theoretical accuracy of $A_{ep}$ and $A_{ed}$ can be improved down to 10 Hz 
by a nonadiabatic calculation of the $(Z\,\alpha)^2\,E_F$ and $(m/M)\,(Z\,\alpha)^2\,E_F$ corrections,
which would allow for more accurate tests of bound-state QED and a more sensitive search for spin-dependent
long-range BSM interactions. 
Moreover, a weighted difference similar to Eq.~(\ref{eq:40}) can be formed for spin-averaged transitions using the extremely
accurate measurements of the $1S$--$2S$ transition in H and D. This would enable a more accurate comparison of theoretical predictions
with measured transitions in H$_2^+$ and HD$^+$, and consequently a significantly more accurate determination of the proton and deuteron masses.


\section*{Data availability}
All data that support the findings of this article are included in the article, in particular in Tables~\ref{tab:I} and \ref{tab:VIII}--\ref{tab:XI}.

\section*{Use of AI tools}
The numerical calculations described in Appendix~\ref{app:num}, including the codes, the figures, 
and parts of the text of Appendix~\ref{app:num}, were prepared with the assistance of the AI system Claude (Anthropic). 
The authors have reviewed the content and take full responsibility for it.

\appendix

\section{Derivation of ${\cal A}^{(7)}_{e\gamma}$}\label{app:A7}
In this Appendix we derive the three parts of ${\cal A}^{(7)}_{e\gamma}$ in Eq.~(\ref{eq:74}) and transform them into the final form presented in Sec.~\ref{sec:A7}.

\subsection{Low-energy contribution}\label{sec:lowen}
In the low-energy region the momentum $k$ of the virtual photon is of the order of $\alpha^2$.
The corresponding low-energy contribution ${\cal A}^{(7)}_L$, 
\begin{align}
{\cal A}^{(7)}_{L} &\ = \frac{2}{3} \int_0^{\lambda} dk\,\tilde f_\gamma(k)\,, \label{eq:A1}
\end{align}
is given by the perturbation of $-f$ in Eq.~(\ref{eq:29}) by ${\cal V}_{\gamma}$,
\begin{align}
{\cal V}_{\gamma}  =&\ \frac{4}{3}\,\pi\,\delta^{(3)}(r_\gamma)\,, \label{eq:A2}
\end{align}
where
\begin{align}
\tilde f_\gamma(k)  =&\ 2 \left< {\cal V}_{\gamma} \frac1{({\cal E}-{\cal H})'}\, \vec{p}\,  \frac{k}{{\cal E}-{\cal H}-k} \, \vec{p} \right>
\nonumber \\ &\
+  \left< \vec{p}\,  \frac{k}{{\cal E}-{\cal H}-k} \,\bigl[ {\cal V}_{\gamma} - \lbr {\cal V}_{\gamma}\rbr\bigr]  \frac1{{\cal E}-{\cal H}-k} \,\vec{p} \right>\,, \label{eq:A3}
\end{align}
and $\vec{p}$ is the electron momentum operator.
The large-$k$ expansion of $\tilde f_\gamma(k)$ is
\begin{align}
\tilde f_\gamma(k) &\ = a_\gamma + \frac{b_\gamma}{\sqrt{k}} + \frac{c_\gamma\,\ln k}{k} + \frac{d_\gamma}{k} + \ldots\,, \label{eq:A4}
\end{align}
where $a_\gamma$, $b_\gamma$, $c_\gamma$, and $d_\gamma$ are the following coefficients:
\begin{align}
a_\gamma =&\
-\frac{2}{3}\, \bigg\langle \vec p^{\;2}\,\frac{1}{({\cal E}-{\cal H})'}\,4\,\pi\,\delta^{(3)}(r_\gamma)\bigg\rangle\,, \label{eq:A5}
\\
b_\gamma =&\  -\frac{16\sqrt{2}}{3}\,\langle \pi\,\delta^{(3)}(r_\gamma)\rangle\,, \label{eq:A6}
 \\
c_\gamma =&\  -\frac{8}{3}\,\langle \pi\,\delta^{(3)}(r_\gamma)\rangle\,, \label{eq:A7}
\end{align}
and
\begin{widetext}
\begin{align}
d_\gamma =&\
\frac{1}{3}\,
\biggl[\langle\vec p\,4\,\pi\,\delta^{(3)}(r_\gamma)\,\vec p\,\rangle
+ 4\,\big\langle 4\,\pi\,\delta^{(3)}(r_\gamma)\big\rangle\,\bigg\langle \frac{1}{r_A}+ \frac{1}{r_B} \bigg\rangle
- 2\,\bigg(\frac{3}{R} + {\cal E}\bigg)\,\big\langle 4\,\pi\,\delta^{(3)}(r_\gamma)\big\rangle
\nonumber \\ &\
+ 2\,\bigg\langle \frac{\vec r_\gamma}{r_\gamma^3}\cdot\bigg(\frac{\vec r_A}{r_A^3}+\frac{\vec r_B}{r_B^3}\bigg)\bigg\rangle
+ \bigg\langle 4\,\pi\,[\delta^{(3)}(r_\gamma)]'\,\frac{1}{({\cal E}-{\cal H})'}\,4\,\pi\,[\delta^{(3)}(r_A) + \delta^{(3)}(r_B)]'\bigg\rangle + 2\,(6-5\,\ln 2)\,\langle4\,\pi\,\delta^{(3)}(r_\gamma)\rangle\biggr]\,. \label{eq:A8}
\end{align}
The $\lambda$-dependent part of ${\cal A}^{(7)}_L$ is separated as
\begin{equation}
{\cal A}^{(7)}_L =  {\cal A}^{(7)}_{L'} + \frac{2}{3}\bigg(a_\gamma\,\lambda + 2\,b_\gamma\,\sqrt{\lambda}
+\frac{c_\gamma}{2}\ln^2\lambda + d_\gamma\,\ln\lambda\bigg)\,. \label{eq:A9}
\end{equation}
Let $t=1/\sqrt{1+2\,k}$, $t_\lambda = 1/\sqrt{1+2\,\lambda}$, and $f_\gamma(t) = \tilde f_\gamma(k)$; then
\begin{align}
{\cal A}^{(7)}_{L'} =&\  \frac{2}{3}\,\lim_{\lambda\rightarrow\infty}
\bigg[ \int_{t_\lambda}^1 dt\,\frac{f_\gamma(t)}{t^3} - \bigg(a_\gamma\,\lambda + 2\,b_\gamma\,\sqrt{\lambda}
+ \frac{c_\gamma}{2}\ln^2\lambda + d_\gamma\,\ln\lambda\bigg)\bigg]
\nonumber \\  =&\
\frac{2}{3}\,\bigg[ \int_0^1 dt\,\frac{f_\gamma(t) - a_\gamma-\sqrt{2}\,b_\gamma\,t+2\,c_\gamma\,\ln(2\,t^2)\,t^2-2\,d_\gamma\,t^2}{t^3}
-\sqrt{2}\, b_\gamma-\frac{c_\gamma}{2}\,\ln^2 2+d_\gamma\,\ln 2 \bigg]\,. \label{eq:A10}
\end{align}
Its numerical calculation is described in Appendix~\ref{app:num}.

\subsection{Second-order contribution}\label{sec:Esec}
 The second-order contribution $E^{(7)}_\mathrm{sec}$ is
\begin{align}
 E^{(7)}_\mathrm{sec} =&\
  2\,\bigg\langle H^{(4)}_{{\rm hfs},A}\frac{1}{({E}-H)'}\,H^{(5)}_A\bigg\rangle
  + 2\,\bigg\langle \kappa\,H^{(4)}_{{\rm hfs},A}\frac{1}{({E}-H)'}\,H^{(4)}_A\bigg\rangle
  + 2\,\bigg\langle H^{(4)}_{{\rm hfs},B}\frac{1}{({E}-H)'}\,2\,\kappa\,H^{(4)}_B\bigg\rangle
   \nonumber \\ =&\
 \sum_\gamma\big[{\cal A}^{(7)}_{e\gamma 1} + {\cal A}^{(7)}_{e\gamma 2}+ {\cal A}^{(7)}_{e\gamma 3}\big]\,\frac{\alpha}{\pi}\,(Z\,\alpha)^6
 \,\frac{m^2}{M_\gamma}\,g_\gamma\,\vec I_\gamma\cdot\vec s_e\,, \label{eq:A11}
\end{align}
where $\kappa\,H^{(4)}_{{\rm hfs},A}$ is the part of Eq.~(\ref{eq:54}) proportional to $\kappa$,
and $H^{(5)}$ is the QED Hamiltonian
\begin{align}
H^{(5)} = &\ H^{(5)}_A + 2\,\kappa\,H^{(4)}_B\,, \label{eq:A12}                 \\
H^{(5)}_A = &\ \bigg(\frac{5}{6} -\frac{1}{5} + \ln\frac{\alpha^{-2}}{2\lambda}\bigg)\,\frac{4\alpha^2 Z}{3m^2}\,\big[\delta^{(3)}(r_A)+\delta^{(3)}(r_B)\big]\,, \label{eq:A13}
\end{align}
where $\kappa = \alpha/(2\,\pi)$. Thus,
\begin{align}
{\cal A}^{(7)}_{e\gamma 1} =&\ 2\,\bigg\langle \frac{4}{3}\,\pi\,\delta^{(3)}(r_\gamma)
\frac{1}{({\cal E}-{\cal H})'}\,\bigg(\frac{5}{6} -\frac{1}{5} + \ln\frac{\alpha^{-2}}{2\lambda}\bigg)\,\frac{4}{3}\,\pi\,\big[\delta^{(3)}(r_A)+\delta^{(3)}(r_B)\big]\bigg\rangle\,, \label{eq:A14}
\\
{\cal A}^{(7)}_{e\gamma 2} =&\ \frac{1}{3}\,\biggl\langle 4\,\pi\,\delta^{(3)}(r_\gamma)
\,\frac{1}{({\cal E}-{\cal H})'}\,\bigg(-\frac{p^4}{8}+\frac{\pi}{2}\,[\delta^{(3)}(r_A) + \delta^{(3)}(r_B)]\bigg)\biggr\rangle = \frac{1}{2}\,{\cal A}^{(6)}_{e\gamma 2}\,, \label{eq:A15}
\\
{\cal A}^{(7)}_{e\gamma 3} =&\ \frac{1}{6}\,\biggl\langle \,\frac{\vec{r}_\gamma}{r_\gamma^3}\times\vec{p}\,\frac{1}{{\cal E}-{\cal H}}\,
\bigg(\frac{\vec{r}_A}{r_A^3} + \frac{\vec{r}_B}{r_B^3}\bigg)\times\vec{p}\biggr\rangle = {\cal A}^{(6)}_{e\gamma 3}\,. \label{eq:A16}
\end{align}
The singularities in these second-order matrix elements cancel with those of the first-order matrix elements.

\subsection{First-order contribution}\label{sec:H7hfs}
The effective Hamiltonian $H^{(7)}_{\rm hfs}$ of order $m\alpha^7$ can be represented as
\begin{align}
\langle H^{(7)}_{\rm hfs} \rangle = &\ \langle H^{(7)}_{{\rm hfs},A} + H^{(7)}_{{\rm hfs},B}
  + \sum_\gamma \zeta_\gamma\,\frac{\alpha}{\pi}\frac{(Z\,\alpha)^3}{m\,M_\gamma}\,g_\gamma\,
  \pi\,\delta^{(3)}(r_\gamma)\,\vec I_\gamma\cdot\vec s_e\rangle
  \nonumber \\ =&\ \sum_\gamma \bigg[A^{(7)}_{e\gamma A} +  A^{(7)}_{e\gamma B} +
  \zeta_\gamma\,\frac{\alpha}{\pi}\frac{(Z\,\alpha)^3}{m\,M_\gamma}\,g_\gamma\,
  \langle \pi\,\delta^{(3)}(r_\gamma)\rangle\bigg]\,\vec I_\gamma\cdot\vec s_e\,, \label{eq:A17}
\end{align}
where $\zeta_\gamma$ accounts for terms that are proportional to $\delta^{(3)}(r_\gamma)$; it is assumed to be independent of $R$
and is obtained by matching with the hydrogenic result.
In the first stage of the derivation we routinely drop such terms; the corresponding
contribution is restored later by matching to the known hydrogenic result; see Sec.~\ref{sec:matching}.
More specifically, we omit terms proportional to $Z^3\,\delta^{(3)}(r_\gamma)$; all other
terms proportional to $\delta^{(3)}(r_\gamma)$ are preserved throughout the derivation.

The first part, $H^{(7)}_{{\rm hfs},A}$,
comes from the spin-dependent terms in the FW Hamiltonian $H_\mathrm{FW}$ \cite{lwqed}
that are proportional to the electron anomalous magnetic moment $\kappa$. Specifically,
\begin{align}
\delta_\kappa H_{\rm FW} = &\ \kappa\,\biggl\{ \sum_\beta \frac{Z\alpha}{2\,m^2}\, \vec\sigma\cdot\frac{\vec r_\beta}{r_\beta^3}\times\vec\pi
+  \frac{e}{8\,m^3}\, \{\vec{\pi}\cdot\vec{B}\,,\, \vec{\pi}\cdot\vec{\sigma} \}
+ \frac{e}{16\,m^3} \,[\pi^i\,,\,[\pi^i\,,\,\vec{\sigma}\cdot \vec{B}]]\biggr\}\,, \label{eq:A18}
\end{align}
so that
\begin{align}
H^{(7)}_{{\rm hfs},A} =&\
\sum_\gamma \kappa \bigg[\sum_\beta \frac{Z\alpha}{2\,m^2}\, \vec\sigma\cdot \frac{\vec r_\beta}{r_\beta^3}\times \big(-e\vec A_\gamma\big)
+\frac{e}{4\,m^3}\,(\vec p\cdot \vec\sigma) (\vec B_\gamma \cdot\vec p)
- \frac{e}{16\,m^3} \, \vec\sigma\cdot\Delta\vec B_\gamma\bigg]\,. \label{eq:A19}
\end{align}
Using $\vec A_\gamma(\vec r)$ and $B_\gamma^i(\vec r)$ given in Eqs.~(\ref{eq:58}) and (\ref{eq:59}), we obtain
\begin{align}
A^{(7)}_{e\gamma A}
= &\ \frac{g_\gamma\,Z\alpha\,\kappa}{2\,M_\gamma}\,\bigg\langle
\frac{2Z\alpha}{3\,m^2}\,\frac{\vec r_\gamma}{r_\gamma^3}\cdot\biggl(\frac{\vec r_A}{r_A^3} + \frac{\vec r_B}{r_B^3}\biggr)
\nonumber \\ &\
-\frac{1}{6\,m^3}\,\bigg(\frac{8\pi}{3}\,p^i\,\delta^{(3)}(r_\gamma)\,p^i
-p^i\,\frac{1}{r_\gamma^5}\big(r_\gamma^2\,\delta^{ij}-3\,r_\gamma^i r_\gamma^j\big)\,p^j\bigg)
+\frac{\pi}{3\,m^3}\,\Delta\,\delta^{(3)}(r_\gamma)
\bigg\rangle\,. \label{eq:A20}
\end{align}
The second part of $H^{(7)}_{\rm hfs}$ is the middle-energy contribution, 
which can be expressed in terms of the slopes of the form factors
and the one-loop vacuum polarization,
\begin{align}
H^{(7)}_{{\rm hfs},B} = &\
 \sum_\gamma -\frac{e}{2\,m} \, \bigg[F'_1(0)+F'_2(0) - \frac{\alpha}{15\,\pi\,m^2}\bigg]\,\vec\sigma \cdot\Delta \, \vec B_\gamma\,, \label{eq:A21}
\end{align}
and thus
\begin{align}
A^{(7)}_{e\gamma B} = &\
\frac{4\pi}{3}\,\frac{g_\gamma\,Z\alpha}{m^3\,M_\gamma}\bigg[
\frac{17}{72} + \frac13\ln\frac{\alpha^{-2}}{2\lambda} -\frac{1}{15}\bigg]\, \frac{\alpha}{\pi}\,\langle\Delta\,\delta^{(3)}(r_\gamma)\rangle\,, \label{eq:A22}
\end{align}
where the slopes of the form factors are given by
\begin{equation}
F'_1(0)+F'_2(0) = \frac{\alpha}{\pi\,m^2}\bigg[\frac{17}{72} + \frac13\ln\frac{\alpha^{-2}}{2\lambda}\bigg]\,. \label{eq:A23}
\end{equation}

\subsection{Elimination of $\lambda$ and divergences from the first-order terms, and matching with hydrogen}\label{sec:matching}
The total result for ${\cal A}^{(7)}_{e\gamma}$ corresponding to Eq.~(\ref{eq:74})  is of the form 
\begin{align}
{\cal A}^{(7)}_{e\gamma} =&\ {\cal A}^{(7)}_L  + {\cal A}^{(7)}_{e\gamma 1} + {\cal A}^{(7)}_{e\gamma 2} + {\cal A}^{(7)}_{e\gamma 3} +
{\cal A}^{(7)}_{e\gamma A} + {\cal A}^{(7)}_{e\gamma B} + \zeta_\gamma\,\langle\pi\delta^{(3)}(r_\gamma)\rangle\,. \label{eq:A24}
\end{align}
The second-order matrix elements ${\cal A}^{(7)}_{e\gamma 1}$ and ${\cal A}^{(7)}_{e\gamma 2}$
contain divergences. They are moved to the first-order matrix elements using Eqs.~(\ref{eq:68}), (\ref{eq:69}), and (\ref{eq:71}). 
Applying these identities, we obtain
\begin{align}
{\cal A}^{(7)}_{e\gamma 1} =&\
\frac{2}{9}\,\bigg(\frac{5}{6} -\frac{1}{5} + \ln\frac{\alpha^{-2}}{2\lambda}\bigg)
\bigg\{\bigg\langle 4\,\pi\,[\delta^{(3)}(r_\gamma)]'\frac{1}{({\cal E}-{\cal H})'}\,4\,\pi\,[\delta^{(3)}(r_A) + \delta^{(3)}(r_B)]'\bigg\rangle
+ 4\,\bigg\langle \bigg(\frac{1}{r_A}+\frac{1}{r_B}\bigg) \bigg\rangle\,\langle 4\,\pi\,\delta^{(3)}(r_\gamma)\rangle
\nonumber \\ &\
- 4\,\bigg\langle \bigg(\frac{1}{r_A}+\frac{1}{r_B}\bigg)\,4\,\pi\,\delta^{(3)}(r_\gamma)\bigg\rangle
+ 2\,\bigg\langle \frac{\vec r_\gamma}{r_\gamma^3}\cdot\bigg(\frac{\vec r_A}{r_A^3}+\frac{\vec r_B}{r_B^3}\bigg)\bigg\rangle\bigg\}\,, \label{eq:A25}
\\
{\cal A}^{(7)}_{e\gamma 2} =&\
\frac{1}{3}\,\bigg\{
\biggl\langle 4\,\pi\,[\delta^{(3)}(r_\gamma)]'\,\frac{1}{({\cal E}-{\cal H})'}\,{\cal H}'^{(4)}_A\biggr\rangle
+ 2\,\bigg\langle \frac{1}{r_\gamma}\bigg\rangle\,{\cal E}^{(4)}
-\langle \pi\,\delta^{(3)}(r_\gamma)\rangle\,\bigg\langle \frac{1}{r_A}+\frac{1}{r_B} \bigg\rangle
\nonumber \\ &\
+ \frac{1}{4}\,\bigg\langle \frac{\vec r_\gamma}{r_\gamma^3}\cdot\bigg(\frac{\vec r_A}{r_A^3}+\frac{\vec r_B}{r_B^3}\bigg)\bigg\rangle
+ \biggl\langle \frac{1}{r_\gamma}\,\bigg({\cal E}+\frac{1}{r_A} + \frac{1}{r_B}\bigg)^2\biggr\rangle
\bigg\}\,. \label{eq:A26}
  \end{align}
The total result is now written in the form
\begin{align}
{\cal A}^{(7)}_{e\gamma} =&\
{\cal A}^{(7)}_{L'} + {\cal A}^{(7)}_\mathrm{sec} + {\cal A}^{(7)}_\mathrm{fo}\,. \label{eq:A27}
\end{align}
The $\ln\lambda$ terms cancel out in ${\cal A}^{(7)}_\mathrm{fo}$,  
and, shifting the terms proportional to $\langle\pi\delta^{(3)}(r_\gamma)\rangle$ to $\zeta_\gamma$, we obtain
\begin{align}
{\cal A}^{(7)}_\mathrm{fo} =&\
\frac{2}{9}\,\bigg(\frac{5}{6} -\frac{1}{5} + \ln\frac{\alpha^{-2}}{2}\bigg)\bigg[
 4\,\bigg\langle \bigg(\frac{1}{r_A}+\frac{1}{r_B}\bigg) \bigg\rangle\,\langle 4\,\pi\,\delta^{(3)}(r_\gamma)\rangle
- \frac{4}{R}\,\bigg\langle 4\,\pi\,\delta^{(3)}(r_\gamma)\bigg\rangle
+ 2\,\bigg\langle \frac{\vec r_\gamma}{r_\gamma^3}\cdot\bigg(\frac{\vec r_A}{r_A^3}+\frac{\vec r_B}{r_B^3}\bigg)\bigg\rangle\bigg]
\nonumber \\ &\
+ \frac{1}{3}\,\bigg[
2\,\bigg\langle \frac{1}{r_\gamma}\bigg\rangle\,{\cal E}^{(4)}
-\langle \pi\,\delta^{(3)}(r_\gamma)\rangle\,\bigg\langle \frac{1}{r_A}+\frac{1}{r_B} \bigg\rangle
+ \frac{1}{4}\,\bigg\langle \frac{\vec r_\gamma}{r_\gamma^3}\cdot\bigg(\frac{\vec r_A}{r_A^3}+\frac{\vec r_B}{r_B^3}\bigg)\bigg\rangle
+ \biggl\langle \frac{1}{r_\gamma}\,\bigg({\cal E}+\frac{1}{r_A} + \frac{1}{r_B}\bigg)^2\biggr\rangle\bigg]
\nonumber \\ &\
+ \frac{1}{4}\,\bigg\langle
\frac{2}{3}\,\frac{\vec r_\gamma}{r_\gamma^3}\cdot\biggl(\frac{\vec r_A}{r_A^3} + \frac{\vec r_B}{r_B^3}\biggr)
-\frac{1}{6}\,\frac{8\pi}{3}\,\vec p\,\delta^{(3)}(r_\gamma)\,\vec p
+\frac{1}{6}\,p^i\,\frac{1}{r_\gamma^5}\big(r_\gamma^2\,\delta^{ij}-3\,r_\gamma^i r_\gamma^j\big)\,p^j
+\frac{\pi}{3}\,\Delta\,\delta^{(3)}(r_\gamma)\bigg\rangle
\nonumber \\ &\
+ \frac{4}{3}\, \bigg[
\frac{17}{72} + \frac13\ln\frac{\alpha^{-2}}{2} -\frac{1}{15}\bigg]\,\langle\pi\,\Delta\,\delta^{(3)}(r_\gamma)\rangle
+ \zeta_\gamma\,\langle\pi\delta^{(3)}(r_\gamma)\rangle\,. \label{eq:A28}
\end{align}
Some operators can be simplified further; the relevant identities, valid up to terms proportional to $\delta^{(3)}(r)$, are
\begin{align}
\langle 4\,\pi\,\Delta \delta^{(3)}(r_\gamma)\rangle =&\
2\,\langle \vec p\,4\,\pi\,\delta^{(3)}(r_\gamma)\,\vec p\rangle -4\,\bigg({\cal E}+\frac{1}{R}\bigg)\langle 4\,\pi\,\delta^{(3)}(r_\gamma)\rangle\,, \label{eq:A29}
\\
\bigg\langle p^i\,\frac{1}{r_\gamma^3}\bigg(\delta^{ij}-3\frac{r_\gamma^i r_\gamma^j}{r_\gamma^2}\bigg)\,p^j\bigg\rangle
=&\  \bigg\langle\frac23\, \vec p\,\pi\,\delta^{(3)}(r_\gamma)\,\vec p +
\frac{\vec r_\gamma}{r_\gamma^3}\cdot\bigg(\frac{\vec r_A}{r_A^3} + \frac{\vec r_B}{r_B^3}\bigg)
-\bigg({\cal E}+\frac{1}{R}\bigg)\,4\,\pi\delta^{(3)}(r_\gamma)\bigg\rangle\,. \label{eq:A30}
\end{align}
\end{widetext}
Using the identities (\ref{eq:A29}) and (\ref{eq:A30})
and matching with hydrogen (see Ref.~\cite{krp:96} and Appendix A of Ref.~\cite{patkos:23}), we obtain ${\cal A}^{(7)}_\mathrm{fo}$ as given in Eq.~(\ref{eq:81}),
with the regularized matrix elements defined in Eqs.~(\ref{eq:76})--(\ref{eq:80}).

\section{Technical details of the numerical calculations for HD$^+$}\label{app:num}
The numerical results for ${\cal E}^{(2)}$, ${\cal E}^{(4)}$, ${\cal E}^{(5)}$,  ${\cal A}^{(4)}_{e\gamma}$, ${\cal A}^{(6)}_{e\gamma}$,
and ${\cal A}^{(7)}_{e\gamma}$ on a predefined grid of $R$ are presented in Table~\ref{tab:VIII}.  
The large-$R$ limits of ${\cal E}^{(i)}$ are equal to the hydrogenic values and are given in
Eqs.~(\ref{eq:13}), (\ref{eq:24}), and (\ref{eq:38}), while 
those of ${\cal A}^{(i)}_{e\gamma}$ are equal to half of the hydrogenic values; see Eqs.~(\ref{eq:15}), (\ref{eq:73}), and (\ref{eq:88}).
The numerical values of all these limits are shown in the last row of Table~\ref{tab:VIII}. 
Below we present technical details of the numerical calculations performed in this work, which exploit the separability of the two-center Schr\"odinger equation.

\subsection{Two-center Schr\"odinger equation}
The two-center Schr\"odinger equation (\ref{eq:10}) is separable in the prolate spheroidal coordinates \cite{hylleraas:31,jaffe:34,baber:35}
\begin{align}
\xi =&\ \frac{r_A+r_B}{R}\,, \label{eq:B1}         \\
\eta =&\ \frac{r_A-r_B}{R}\,. \label{eq:B2}
\end{align}
Consequently, the wave function of the ground electronic state is of the form
\begin{align}
\phi(\xi,\eta) = \Lambda(\xi)\,M(\eta)\,, \label{eq:B3}
\end{align}
where $\Lambda$ and $M$ satisfy the following differential equations
\begin{align}
\frac{d}{d\xi}\biggl[(\xi^2-1)\,\frac{d\Lambda}{d\xi}\biggr] + \bigl(A + 2\,R\,\xi - p^2\,\xi^2\bigr)\,\Lambda =&\ 0\,, \label{eq:B4}         \\
\frac{d}{d\eta}\biggl[(1-\eta^2)\,\frac{dM}{d\eta}\biggr] + \bigl(-A + p^2\,\eta^2\bigr)\,M =&\ 0\,, \label{eq:B5}
\end{align}
$A$ is the separation constant, and
\begin{align}
p^2 = -\frac{{\cal E}\,R^2}{2}\,. \label{eq:B6}
\end{align}
The function $M$ is expanded in the Legendre polynomials with even $l$,
\begin{align}
M(\eta) = \sum_l c_l\,P_l(\eta)\,, \label{eq:B7}
\end{align}
and for $\Lambda$ we use the expansion of Jaff\'e \cite{jaffe:34},
\begin{align}
\Lambda(\xi) =&\ e^{-p\,\xi}\,(\xi+1)^{R/p-1}\,\sum_n g_n\,\biggl(\frac{\xi-1}{\xi+1}\biggr)^n\,. \label{eq:B8}
\end{align}
The coefficients $c_l$ and $g_n$ satisfy three-term recurrence relations, and the requirement that both of them have 
solutions with the same $A$ determines $p$, and thus the energy ${\cal E}(R)$ \cite{bates:53,madsen:71}. 
We solve these equations in arbitrary-precision arithmetic, 
so that ${\cal E}(R)$ and the wave function are obtained with more than 30 significant digits. 
As a check, our result at the equilibrium distance $R=2$ is in perfect agreement 
with the result of Ref.~\cite{fernandez:21}, which is accurate to more than 100 digits.

\subsection{First-order matrix elements}
The first-order matrix elements in ${\cal E}^{(4)}$, ${\cal E}^{(5)}$, ${\cal A}^{(6)}_{e\gamma}$, and ${\cal A}^{(7)}_\mathrm{fo}$, 
including the regularized ones in Eqs.~(\ref{eq:76}) and (\ref{eq:77}), 
are calculated by direct numerical integration with this exact wave function.
In the spheroidal coordinates of Eqs.~(\ref{eq:B1}) and (\ref{eq:B2})
\begin{align}
r_A =&\ \frac{R}{2}\,(\xi+\eta)\,, \label{eq:B9}     \\
r_B =&\ \frac{R}{2}\,(\xi-\eta)\,, \label{eq:B10}     \\
d^3r =&\ \frac{R^3}{8}\,(\xi^2-\eta^2)\,d\xi\,d\eta\,d\varphi\,, \label{eq:B11}
\end{align}
so that these matrix elements are two-dimensional integrals over $\xi\in(1,\infty)$ and $\eta\in(-1,1)$.
The differential operators are eliminated beforehand with the help of the Schr\"odinger equation (\ref{eq:10}), for example
\begin{align}
\langle p^4\rangle = 4\,\biggl\langle\biggl({\cal E}+\frac{1}{r_A}+\frac{1}{r_B}\biggr)^2\biggr\rangle\,, \label{eq:B12}
\end{align}
and the expectation value of the Dirac $\delta$ function, for $\phi$ normalized to unity, is
\begin{align}
\langle \delta^{(3)}(r_\gamma)\rangle =  \phi^2(1, \eta_\gamma)\,, \label{eq:B13}
\end{align}
where $\eta_A=-1$ and $\eta_B = 1$.
In ${\cal E}^{(4)}$ and ${\cal A}^{(6)}_{e\gamma}$ the factors $(\xi\pm\eta)^{-1}$ are expanded according to Eq.~(\ref{eq:B28}) below. 
The integral over $\eta$ is then performed analytically with the Legendre expansion of the $\eta$-dependent part of the integrand, 
and the remaining integral over $\xi$ is evaluated by an adaptive quadrature in arbitrary-precision arithmetic.
In ${\cal A}^{(7)}_\mathrm{fo}$ the matrix elements with integrable singularities, $\langle 1/(r_A\,r_B)\rangle$, $\langle 1/(r_A^2\,r_B)\rangle$, 
and $\langle \vec r_A\cdot\vec r_B/(r_A^3\,r_B^3)\rangle$, are evaluated with the Gauss--Legendre quadrature in $\xi$ and $\eta$ 
on intervals graded geometrically towards the nuclei. 
For the regularized matrix elements in Eqs.~(\ref{eq:76}) and (\ref{eq:77}) with $\gamma=A$ we use the variables $u=\xi+\eta=2\,r_A/R$ and $v=\xi-\eta$, 
and first integrate over $v$,
\begin{align}
g(u) = \int dv\,v\,\phi^2\,, \label{eq:B14}
\end{align}
where $|2-u|<v<2+u$. The small-$u$ expansion of this function is
\begin{align}
g(u) = g_1\,u + g_2\,u^2 + O(u^3)\,, \label{eq:B15}
\end{align}
where $g_1 = 4\,\phi^2$ at the nucleus $A$ and $g_2 = -R\,g_1$ follows from the cusp condition. 
These two terms are subtracted from $g(u)$ and integrated analytically with $u^{-2}$ and $u^{-3}$, 
which gives the $\epsilon\rightarrow0$ limit in Eqs.~(\ref{eq:76}) and (\ref{eq:77}), 
while the remainder is integrated numerically.
Finally, the matrix element in Eq.~(\ref{eq:80}) is obtained from the nonspherical part of the gradient of $\phi$ at the nucleus,
\begin{align}
\langle \vec p\;\delta^{(3)}(r_A)\,\vec p\,\rangle =&\ \frac{1}{R^2}\,
\biggl(\Lambda\,\frac{dM}{d\eta} - M\,\frac{d\Lambda}{d\xi}\biggr)^2\,\bigg|_{\xi=1,\,\eta=-1}\,, \label{eq:B16}
\end{align}
and it vanishes in the hydrogenic limit. 
As a test, the normalization of $\phi$, $\langle 1/r_A\rangle$, and $\langle 1/r_A^2\rangle$ obtained with these different integration schemes 
agree to better than $10^{-15}$ at all distances $R$.
In comparison with previous calculations, the mass--velocity and Darwin terms of Ref.~\cite{tsogbayar:06} give
${\cal E}^{(4)}$ within $8.1\times10^{-11}$ of our values at the 13 common
distances. 

\subsection{Bethe logarithm}
The calculation of the Bethe logarithm requires the solution of the first-order equation
\begin{align}
({\cal H}-{\cal E}+k)\,\vec\psi = \vec\nabla\phi \label{eq:B17}
\end{align}
for all values of the photon momentum $k$. 
We again take advantage of the separability: for functions with the angular 
dependence $e^{\pm i\,m\,\varphi}$  around the internuclear axis, the operator
\begin{align}
\frac{R^2}{2}\,(\xi^2-\eta^2)\,({\cal H}-{\cal E}+k) = {\cal T}_\xi + {\cal T}_\eta\,, \label{eq:B18}
\end{align}
is a sum of an operator in $\xi$ and an operator in $\eta$,
\begin{align}
{\cal T}_\xi =&\ -\frac{\partial}{\partial\xi}\,(\xi^2-1)\,\frac{\partial}{\partial\xi} + \frac{m^2}{\xi^2-1} - 2\,R\,\xi + q^2\,\xi^2\,, \label{eq:B19}         \\
{\cal T}_\eta =&\ -\frac{\partial}{\partial\eta}\,(1-\eta^2)\,\frac{\partial}{\partial\eta} + \frac{m^2}{1-\eta^2} - q^2\,\eta^2\,, \label{eq:B20}
\end{align}
where
\begin{align}
q^2 = \frac{(k-{\cal E})\,R^2}{2}\,. \label{eq:B21}
\end{align}
Therefore, Eq.~(\ref{eq:B17}) is solved in the product basis 
\begin{align}
\chi_{nl}^m(\xi,\eta) =&\ (\xi^2-1)^{m/2}\,e^{-\beta\,(\xi-1)}
\nonumber \\ &\times L^{(m)}_n\bigl(2\,\beta\,(\xi-1)\bigr)\,P^m_l(\eta)\,, \label{eq:B22}
\end{align}
where $L^{(m)}_n$ are the associated Laguerre polynomials, $P^m_l$ are the associated Legendre functions, $m=0,1$, and
\begin{align}
\beta = \sqrt{p\,q}\,. \label{eq:B23}
\end{align}
The right-hand side of Eq.~(\ref{eq:B17}) is calculated with the exact wave function $\phi$ of Eq.~(\ref{eq:B3}), and its projections on the basis functions
are obtained by the Gauss--Legendre quadrature in $\eta$ and the Gauss--Laguerre quadrature in $\xi$.
Since $(\xi^2-\eta^2)\,\vec\nabla\phi$ consists of two products of functions of $\xi$ and of $\eta$, and the solution is analytic in both variables,
the linear system reduces to a tridiagonal eigenvalue problem in $l$ and banded systems in $n$, and the expansion converges exponentially.

Using quadruple-precision arithmetic, $f(t)$ of Eq.~(\ref{eq:29}) is obtained with $25$--$30$ significant digits. 
The integral over $t$  in Eq.~(\ref{eq:27}) is evaluated with the Gauss--Legendre quadrature on intervals graded towards $t=0$ down to $t_0\sim 10^{-3}$.
For $t<t_0$ we use the expansion in Eq.~(\ref{eq:30}), supplemented by terms $t^n$ and $t^n\ln t$ up to $n=12$, with coefficients fitted 
to the calculated values. The numerical fit reproduces $f_3$ and $f_4$ of Eqs.~(\ref{eq:33}) and (\ref{eq:34}) to better than $10^{-10}$,
which gives us confidence in the accuracy of $f(t)$ for $t>t_0$.

One more factor has to be taken into account. 
At large $R$ the state $1s\sigma_u$ with the wave function $\phi_u$ is nearly degenerate with the ground state. Its contribution to $f$,
\begin{align}
f_u(t) =&\ |\langle\phi_u|\vec p\,|\phi\rangle|^2\,\frac{k}{k+\Delta}\,, \label{eq:B24}
\end{align}
where $\Delta$ is the excitation energy, varies rapidly near $t=1$. It is therefore subtracted from $f(t)$,
and its contribution to the Bethe logarithm,
\begin{align}
D\,\ln k_0\bigl|_u = |\langle\phi_u|\vec p\,|\phi\rangle|^2\,\Delta\,\ln(2\,\Delta)\,, \label{eq:B25}
\end{align}
is added analytically.
We observe moderate agreement with the previous calculations of Ref.~\cite{korobov:13}.
For example, Eq.~(\ref{eq:25}) with our contact term and $\ln k_0$ from the Supplemental Material of Ref.~\cite{korobov:13}
differs from our ${\cal E}^{(5)}$ by $2.6\times10^{-5}$, $4.4\times10^{-10}$, and $9.2\times10^{-7}$
at $R=0.1$, $2$, and $7$, respectively, while all our results are accurate to $10^{-12}$.

\subsection{Second-order matrix elements}
In the second-order terms of ${\cal A}^{(6)}_{e\gamma}$ in Eq.~(\ref{eq:72}) and of ${\cal A}^{(7)}_{e\gamma}$ in Eq.~(\ref{eq:83}), 
both $4\,\pi\,[\delta^{(3)}(r_\gamma)]'\phi$ and ${\cal H}'^{(4)}_A\phi$ behave as $r_\gamma^{-2}$ at the nucleus.
Thus, the first-order wave function contains $\ln r_\gamma$, which cannot be represented efficiently by polynomials in $\xi$ and $\eta$. 
Therefore, we use the fact that $(\xi^2-\eta^2)\,4\,\pi\,[\delta^{(3)}(r_A)]'\phi$ is equal to $(\xi+\eta)^{-2}$ 
times a sum of two products of functions of $\xi$ and of $\eta$, and employ the integral representation
\begin{align}
\frac{1}{(\xi+\eta)^{2}} = \int_0^\infty ds\,s\,e^{-s\,(\xi+\eta)}\,. \label{eq:B26}
\end{align}
The second-order matrix element with an arbitrary operator $Q$ then becomes
\begin{align}
\bigg\langle 4\,\pi\,[\delta^{(3)}(r_A)]'\,\frac{1}{({\cal E}-{\cal H})'}\,Q\bigg\rangle = \int_0^\infty ds\,g(s)\,, \label{eq:B27}
\end{align}
where $g(s)$ involves the solution of the first-order equation with a source that is analytic in $\xi$ and $\eta$. 
This equation is solved in the product basis of Eq.~(\ref{eq:B22}) with $k=0$ and with a single exponent for all values of $s$,
chosen arbitrarily to be $ \beta = \sqrt{2\,p\,s_\mathrm{max}}$, 
where $s_\mathrm{max}$ is the upper limit of the numerical integration over $s$ and is a few hundred.
Because the operator ${\cal H}-{\cal E}$ does not depend on $s$, the matrices and the projections of $Q\,\phi$ are calculated only once for each $R$, 
while the number of basis functions in $\xi$ is increased to achieve high numerical precision. 

The projections of $Q\,\phi$, namely of ${\cal H}'^{(4)}_A\phi$ or $(\vec r_A/r_A^3+\vec r_B/r_B^3)\times\vec p\,\phi$, on the basis functions are obtained using the expansion
\begin{align}
\frac{1}{\xi\pm\eta} = \sum_{l=0}^\infty (2\,l+1)\,(\mp1)^l\,Q_l(\xi)\,P_l(\eta)\,, \label{eq:B28}
\end{align}
where $Q_l$ are the Legendre functions of the second kind (the terms with $(\xi\pm\eta)^{-2}$ are reduced to this form by integration by parts).
The function $g(s)$ is calculated for $s\leq s_\mathrm{max}$, and the integral over $s>s_\mathrm{max}$ is obtained from the expansion
\begin{align}
g(s) = \sum_{n\geq2} \frac{a_n + b_n\,\ln s}{s^{n}}\,, \label{eq:B29}
\end{align}
with the coefficients fitted to the calculated values.  

The last term in Eq.~(\ref{eq:72}) is calculated in the same way. The operator $\vec r_\gamma/r_\gamma^3\times\vec p$ 
changes the symmetry of the wave function from $\sigma$ to $\pi$, so that the first-order wave function is expanded in the basis functions $\chi^1_{nl}$ of Eq.~(\ref{eq:B22}),
and the resolvent does not require the subtraction of the ground state. For $\gamma=A$ we have
\begin{align}
(\xi^2-\eta^2)\,\frac{\vec r_A}{r_A^3}\times\vec\nabla\phi =&\ \frac{8}{R^3}\,\frac{\sqrt{(\xi^2-1)\,(1-\eta^2)}}{(\xi+\eta)^2}
\nonumber \\ &\ \times
\biggl(\frac{d\Lambda}{d\xi}\,M - \Lambda\,\frac{dM}{d\eta}\biggr)\,\vec e_\varphi\,, \label{eq:B30}
\end{align}
where $\vec e_\varphi$ is the unit vector in the azimuthal direction around the internuclear axis, and the result for $\gamma=B$ is obtained by the replacement $\eta\rightarrow-\eta$.
This source is again $(\xi+\eta)^{-2}$ times a sum of two products of functions of $\xi$ and of $\eta$, and it behaves as $r_A^{-2}$ at the nucleus.
We therefore use the integral representation in Eq.~(\ref{eq:B26}), and for each $s$ solve the first-order equation with an analytic source 
in the same basis, with the same $\beta$ and $s_\mathrm{max}$ as above.
The projections of $(\vec r_A/r_A^3+\vec r_B/r_B^3)\times\vec p\,\phi$ on the basis functions $\chi^1_{nl}$ are obtained with the help of Eq.~(\ref{eq:B28}),
and the integral over $s>s_\mathrm{max}$ from the expansion in Eq.~(\ref{eq:B29}).

\subsection{Scalar hfs correction to the Bethe logarithm}
The scalar hfs correction to the Bethe logarithm is calculated using
the same product basis of Eq.~(\ref{eq:B22}) with $\beta$ from Eq.~(\ref{eq:B23}) as for the ordinary Bethe logarithm.
The function $\tilde f_\gamma(k)$ of Eq.~(\ref{eq:A3}) is expressed 
in terms of the solution $\vec\psi$ of Eq.~(\ref{eq:B17}) as
\begin{align}
\tilde f_\gamma(k) =&\ k\,\Bigl[2\,\langle\phi|{\cal V}_\gamma|\psi\rangle + \langle\vec\psi\,|{\cal V}_\gamma - \langle{\cal V}_\gamma\rangle|\vec\psi\,\rangle\Bigr]\,, \label{eq:B31}         \\
\psi =&\ \frac{1}{({\cal E}-{\cal H})'}\,\vec\nabla\cdot\vec\psi\,, \label{eq:B32}
\end{align}
where $\phi$ is normalized to unity. Since $(\xi^2-\eta^2)\,\vec\nabla\cdot\vec\psi$ is analytic in $\xi$ and $\eta$, 
the function $\psi$ is expanded in the basis functions $\chi^0_{nl}$ of Eq.~(\ref{eq:B22}) with even $l$ and with the same exponent $\beta$ of Eq.~(\ref{eq:B23}) 
as used for $\vec\psi$ at the given $k$. The equation for $\psi$ is solved using Eq.~(\ref{eq:B18}) at $k=0$, and the reduced resolvent is obtained 
by projecting out the ground state represented in this basis.
Because ${\cal V}_\gamma$ of Eq.~(\ref{eq:A2}) is proportional to $\delta^{(3)}(r_\gamma)$,
\begin{align}
\langle\phi|{\cal V}_\gamma|\psi\rangle =&\ \frac{4\,\pi}{3}\,\phi(1,\eta_\gamma)\,\psi(1,\eta_\gamma)\,, \label{eq:B33}        \\
\langle\vec\psi\,|{\cal V}_\gamma|\vec\psi\,\rangle =&\ \frac{4\,\pi}{3}\,|\vec\psi(1,\eta_\gamma)|^2\,, \label{eq:B34}
\end{align}
as in Eq.~(\ref{eq:B13}). These values are calculated directly from the expansions in the basis functions.
No regularization of ${\cal V}_\gamma$ is needed here, because the functions $\phi$, $\psi$, and $\vec\psi$ have only Coulomb cusps at the nuclei 
and are therefore analytic in $\xi$ and $\eta$.
The integral in Eq.~(\ref{eq:A10}) is evaluated with the Gauss--Legendre quadrature on intervals graded towards $t=0$ and $t=1$, down to
$t_0\sim 10^{-3}$--$10^{-2}$, and for $t<t_0$ the integrand is represented by the terms $t^n\ln^j t$ with $n\leq4$ and $j\leq2$, fitted to the calculated values.

\subsection{DVR averaging}
With all the potentials of Table~\ref{tab:VIII} at hand, the radial equation for $\chi_{vL}$ is solved with 
the discrete variable representation of Colbert and Miller \cite{colbert:92} and with the CODATA 2022 masses of Table~\ref{tab:VI}.
The potentials ${\cal E}^{(2)}(R)$, ${\cal A}^{(4)}_{e\gamma}(R)$, and ${\cal E}^{(4)}(R)$ are evaluated exactly on demand
for the DVR averaging, while for ${\cal A}^{(6)}_{e\gamma}$, ${\cal A}^{(7)}_{e\gamma}$, and ${\cal E}^{(5)}$ the smooth
ratios ${\cal A}^{(6)}_{e\gamma}/{\cal A}^{(4)}_{e\gamma}$, ${\cal A}^{(7)}_{e\gamma}/{\cal A}^{(4)}_{e\gamma}$, and ${\cal E}^{(5)}/{\cal A}^{(4)}_{e\gamma}$ are interpolated by splines in $\ln R$.
With the DVR eigenvector components $\chi_{vL}(R_k)$ normalized as $\sum_k\chi^2_{vL}(R_k)=1$, the first-order averages are 
\begin{align}
\Delta^{(i)}_{vL}=\sum_k\chi^2_{vL}(R_k)\,\tilde {\cal A}^{(i)}_{e\gamma}(R_k)\,, \label{eq:B35}
\end{align}
and the second-order ones
\begin{align}
\delta^{(i)}_{vL}=2\sum_{n\neq v}
\frac{\langle v,L|{\cal E}^{(i-2)}|n,L\rangle\langle n,L|{\cal A}^{(4)}_{e\gamma}|v,L\rangle}{{\cal E}_{vL}-{\cal E}_{nL}}\,, \label{eq:B36}
\end{align}
where the sum over the complete DVR spectrum for the same $L$ represents the reduced resolvent;
for $i=7$ the term $\delta^{(6)}_{vL}/2$ of Eq.~(\ref{eq:46}) is added.
The averages are given in Table~\ref{tab:IX}. $\delta$ and $\Delta$ are of the same order and tend to cancel in the sum.

Using the sums $\delta+\Delta$, we obtain $A^{(6)}_{e\gamma}$ and $A^{(7)}_{e\gamma}$ in Hz, which are presented
in Tables~\ref{tab:X} ($\gamma=p$) and \ref{tab:XI} ($\gamma=d$). 
For the ground state $A^{(6)}_{ep}=-7672.28$~Hz and $A^{(7)}_{ep}=-538.32$~Hz,
of which $+17.68$~Hz comes from the anomalous magnetic moment. 
We note that $A^{(7)}_{ep}/A^{(6)}_{ep} = 0.07$ is much larger than the naive estimate of 
a single power of the fine-structure constant $\alpha$. 
The dependence on $v$ is a few percent per vibrational quantum, and
that on $L$ is about one percent over $L=0$--$3$.

\subsection{Nonadiabatic contact term}
Tables~\ref{tab:X} ($\gamma=p$) and \ref{tab:XI} ($\gamma=d$) list the nonadiabatic expectation values
$\langle\pi\delta^{(3)}(r_{e\gamma})\rangle_{vL}$, obtained with the full three-body wave function \cite{pantak:26} and the mass ratios of
Table~\ref{tab:VI}, the first term of Eq.~(\ref{eq:42}), $\bar A_{ep}$ and $\bar A_{ed}$, and the totals $A_{ep}$ and $A_{ed}$, which are
collected by $v$ and $L$ in Table~\ref{tab:I}. 
For the ground state, the nonadiabatic $\langle\pi\delta^{(3)}(r_{ep})\rangle_{00}=0.650\,444$
lies $0.18\%$ below the Born--Oppenheimer value $0.651\,633$, which is the expected finite-nuclear-mass effect.
For $v\leq4$ and $L\leq3$, after accounting for the nuclear masses, the values of $\langle\pi\delta^{(3)}(r_{e\gamma})\rangle_{vL}$ 
in Tables~\ref{tab:X} and \ref{tab:XI} agree, within the numerical precision quoted in these tables,
with the results of Zhong {\em et al.} from Table~II of Ref.~\cite{zhong:12}.

The comparison with the earlier results of Aznabayev \emph{et al.} \cite{Korobov:19} is more complicated.
The values of $\langle\pi\delta^{(3)}(r_{e\gamma})\rangle_{vL}$ tabulated in Table~III of Ref.~\cite{Korobov:19} 
are systematically higher than ours. For example, for $5\leq v\leq9$ and $L\leq3$, all 40 of their values are higher than ours,
after accounting for the nuclear masses, by up to $3.4\times10^{-10}$.

\begin{table*}[htb]
\caption{\label{tab:VIII} Nonrelativistic energy ${\cal E}^{(2)}$ of Eq.~(\ref{eq:12}), the leading relativistic correction ${\cal E}^{(4)}$ of Eq.~(\ref{eq:22}),
and the leading QED correction ${\cal E}^{(5)}$ of Eq.~(\ref{eq:25}) of the ground electronic state of HD$^+$ in the Born--Oppenheimer approximation, 
as functions of the internuclear distance $R$ (atomic units). 
The last three columns are the contact spin interaction coefficients ${\cal A}^{(4)}_{e\gamma}$ and 
${\cal A}^{(6)}_{e\gamma}$, and  ${\cal A}^{(7)}_{e\gamma}$ of Eq.~(\ref{eq:A27}).
All digits of ${\cal E}^{(2)}$, ${\cal E}^{(4)}$,  ${\cal E}^{(5)}$, ${\cal A}^{(4)}_{e\gamma}$ and 
${\cal A}^{(6)}_{e\gamma}$  are are numerically converged to the displayed precision, while
the numerical uncertainty of   ${\cal A}^{(7)}_{e\gamma}$ comes from the small-$t$ extrapolation of ${\cal A}^{(7)}_{L'}$, and
is given in parentheses when it exceeds one half of the last digit. The last row gives the exact asymptotic values from the analytic formulas.}
\footnotesize
\begin{ruledtabular}
\begin{tabular}{d{2.2}d{2.16}d{2.16}d{3.14}d{1.16}d{2.12}d{5.9}}
\centt{$R$} & \centt{${\cal E}^{(2)}(R)$} & \centt{${\cal E}^{(4)}(R)$} & \centt{${\cal E}^{(5)}(R)$} & \centt{${\cal A}^{(4)}_{e\gamma}(R)$} & \centt{${\cal A}^{(6)}_{e\gamma}(R)$} & \centt{${\cal A}^{(7)}_{e\gamma}(R)$}\\ \hline
0.1 & 8.0217579291522481 & -1.7689628599497096 & 113.980903683862 & 8.4925875070351153 & 32.282317746656 & -2159.331120(7)\\
0.2 & 3.0713796982966413 & -1.4389128813468541 & 93.237019490223 & 6.7090737852806513 & 17.956254346615 & -969.109763(5)\\
0.3 & 1.4666292539032291 & -1.1531212196195657 & 76.233151651364 & 5.3740114727477430 & 10.758901035474 & -513.745819(2)\\
0.4 & 0.6992459405698327 & -0.9284619818303230 & 63.056063594903 & 4.3852492245084879 & 6.939732097363 & -302.348329(1)\\
0.5 & 0.2650120000264269 & -0.7562294075989127 & 52.921345203861 & 3.6457162471019745 & 4.795335047574 & -193.766061(1)\\
0.6 & -0.0048180477784655 & -0.6244380044284190 & 45.071832930313 & 3.0837513430682842 & 3.524699675894 & -133.843674\\
0.7 & -0.1826248370438247 & -0.5229004039104137 & 38.920414399693 & 2.6494027212250304 & 2.733618788643 & -98.827418\\
0.8 & -0.3044800944453221 & -0.4438780133203365 & 34.036969643646 & 2.3081601205559353 & 2.218566725313 & -77.355878\\
0.9 & -0.3902704881246756 & -0.3816957789992505 & 30.110533760544 & 2.0359966111689939 & 1.869514983712 & -63.628823\\
1.0 & -0.4517863133784512 & -0.3322267472399534 & 26.915685368453 & 1.8159505224078085 & 1.624384812110 & -54.524635\\
1.1 & -0.4964118670566773 & -0.2924614297663106 & 24.287517317734 & 1.6358562827073386 & 1.446752740281 & -48.286902\\
1.2 & -0.5289745245003295 & -0.2601881550703287 & 22.103974285603 & 1.4868465443408775 & 1.314469939460 & -43.888645\\
1.3 & -0.5527406218804872 & -0.2337653529383328 & 20.273582512024 & 1.3623565544138509 & 1.213620062930 & -40.708726\\
1.4 & -0.5699835280567854 & -0.2119614433145061 & 18.726927930729 & 1.2574542185093965 & 1.135192991157 & -38.359933\\
1.5 & -0.5823232054549776 & -0.1938423480866308 & 17.410698323658 & 1.1683828059257004 & 1.073191740607 & -36.594171\\
1.6 & -0.5909372245053716 & -0.1786920546438762 & 16.283477050703 & 1.0922441089714867 & 1.023521617378 & -35.248436\\
1.7 & -0.5966962695398744 & -0.1659560662689392 & 15.312740050373 & 1.0267756502067330 & 0.983319814860 & -34.213096\\
1.8 & -0.6002536341023100 & -0.1552007632178995 & 14.472686849069 & 0.9701917719408994 & 0.950540204424 & -33.412785\\
1.85 & -0.6013667165915524 & -0.1504559829600277 & 14.095043429824 & 0.9447749634586318 & 0.936450722377 & -33.083578\\
1.9 & -0.6021057828832493 & -0.1460839038314011 & 13.742655717942 & 0.9210687416741430 & 0.923689939413 & -32.794592\\
1.95 & -0.6025152395971052 & -0.1420524803242756 & 13.413552079363 & 0.8989379084225687 & 0.912129750559 & -32.541506\\
2.0 & -0.6026342144949465 & -0.1383329938679796 & 13.105952528641 & 0.8782606089882357 & 0.901658603344 & -32.320623\\
2.05 & -0.6024972590579457 & -0.1348996563190293 & 12.818246910199 & 0.8589267856585478 & 0.892178936591 & -32.128767\\
2.1 & -0.6021349458445536 & -0.1317292660982801 & 12.548975886334 & 0.8408368343788357 & 0.883605063585 & -31.963196\\
2.15 & -0.6015743781925450 & -0.1288009092983850 & 12.296814331017 & 0.8239004347982448 & 0.875861414794 & -31.821530\\
2.2 & -0.6008396265944108 & -0.1260956992953139 & 12.060556791007 & 0.8080355287572398 & 0.868881072814 & -31.701696\\
2.3 & -0.5989308861578464 & -0.1212879739615258 & 11.631455297939 & 0.7792280207840891 & 0.856978736926 & -31.520469\\
2.4 & -0.5965536385920330 & -0.1171875835531707 & 11.253973258434 & 0.7538917377506212 & 0.847493772064 & -31.407396\\
2.5 & -0.5938235109904680 & -0.1136965424839047 & 10.921653732101 & 0.7315892533648567 & 0.840099390392 & -31.352902\\
2.6 & -0.5908331963248127 & -0.1107332876240208 & 10.629045721392 & 0.7119519565253180 & 0.834527459779 & -31.349271\\
2.7 & -0.5876572896786272 & -0.1082294795765374 & 10.371518580138 & 0.6946672159694138 & 0.830554720518 & -31.390150\\
2.8 & -0.5843560034674206 & -0.1061274879451224 & 10.145114520878 & 0.6794682137502226 & 0.827992571577 & -31.470204\\
2.9 & -0.5809780452697365 & -0.1043784007381149 & 9.946430583729 & 0.6661258338655831 & 0.826679413914 & -31.584857\\
3.0 & -0.5775628640489736 & -0.1029404385386336 & 9.772523568272 & 0.6544421466411239 & 0.826474852295 & -31.730102\\
3.1 & -0.5741424157159233 & -0.1017776836220909 & 9.620833000562 & 0.6442451420810367 & 0.827255267496 & -31.902365\\
3.2 & -0.5707425598956967 & -0.1008590559447627 & 9.489118370492 & 0.6353844483363284 & 0.828910413518 & -32.098404\\
3.35 & -0.5657259108970165 & -0.0998805617232138 & 9.324750580732 & 0.6243139746099997 & 0.832818031494 & -32.430554\\
3.5 & -0.5608555387980381 & -0.0993133440772031 & 9.195207448566 & 0.6155718791311381 & 0.838171305740 & -32.800214\\
3.75 & -0.5531617014029679 & -0.0991103106804599 & 9.043871774825 & 0.6053181015397773 & 0.849591954672 & -33.476629\\
4.0 & -0.5460848837129383 & -0.0996305963763857 & 8.957779479717 & 0.5994258055752282 & 0.863209692453 & -34.197227\\
4.25 & -0.5396743135638569 & -0.1006837459936110 & 8.922389033618 & 0.5969239264892539 & 0.878089482493 & -34.929266\\
4.5 & -0.5339400311092339 & -0.1021158201360817 & 8.925856861495 & 0.5970222359432166 & 0.893428753193 & -35.643926\\
4.75 & -0.5288662614029215 & -0.1037999405796949 & 8.958356912716 & 0.5990654979919224 & 0.908556633143 & -36.317036\\
5.0 & -0.5244202951676077 & -0.1056309143413121 & 9.011669401519 & 0.6025057690017345 & 0.922939882728 & -36.929863\\
5.5 & -0.5172315078578577 & -0.1094044936716679 & 9.154507909045 & 0.6118293345439856 & 0.948045688895 & -37.929491\\
6.0 & -0.5119690484666523 & -0.1129409477807139 & 9.313388965963 & 0.6222602355398188 & 0.967204368357 & -38.608906\\
7.0 & -0.5055940042392590 & -0.1184130822016879 & 9.592044680022 & 0.6406161836030249 & 0.989178837117 & -39.190958\\
8.0 & -0.5025703885999690 & -0.1216660854064982 & 9.775367253696 & 0.6527177338096528 & 0.997345149253 & -39.215695\\
9.0 & -0.5011954528983677 & -0.1233668234219922 & 9.878168978140 & 0.6595116290844090 & 0.999703634642 & -39.087594\\
10.0 & -0.5005787289439264 & -0.1241981389540587 & 9.931301955458 & 0.6630259155785928 & 1.000161993028 & -38.965701\\
11.0 & -0.5002992293703975 & -0.1245932668988955 & 9.957738695680 & 0.6647757399967498 & 1.000138659097 & -38.883432\\
12.0 & -0.5001683069678789 & -0.1247811608623925 & 9.970769026369 & 0.6656388316787986 & 1.000051483682 & -38.834821\\
\centt{$\infty$} & -0.5 & -0.125 & 9.98625612749328 & 0.6666666666666667 & 1 & -38.774038755\\
\end{tabular}
\end{ruledtabular}
\end{table*}

\begin{table*}[htb]
\caption{\label{tab:IX} The rovibrational averages $\delta$ and $\Delta$ defined in Eqs.~(\ref{eq:43})--(\ref{eq:46}). 
$\delta^{(6)}_{vL}$ is converged to all digits shown. The uncertainties of $\Delta^{(6)}_{vL}$ and $\delta^{(7)}_{vL}$ ($\sim10^{-9}$) come from the interpolation of ${\cal A}^{(6)}_{e\gamma}(R)$ and ${\cal E}^{(5)}(R)$ between the grid points of Table~\ref{tab:VIII}, and that of $\Delta^{(7)}_{vL}$ ($\sim10^{-6}$) from the numerical uncertainty of ${\cal A}^{(7)}_{e\gamma}(R)$.}

\begin{ruledtabular}
\begin{tabular}{mmd{1.10}d{2.10}d{2.10}d{3.7}d{2.6}d{2.6}}
\centt{$v$} & \centt{$L$} & \centt{$\delta^{(6)}_{vL}$} & \centt{$\Delta^{(6)}_{vL}$} & \centt{$\Delta^{(6)}_{vL}+\delta^{(6)}_{vL}$} & \centt{$\delta^{(7)}_{vL}$} & \centt{$\Delta^{(7)}_{vL}$} & \centt{$\Delta^{(7)}_{vL}+\delta^{(7)}_{vL}$}\\ \hline
0 & 0 & 0.2681776981 & -0.4033505589 & -0.1351728608 & -22.2530170 & 18.169944 & -4.083073\\
0 & 1 & 0.2671480044 & -0.4025706898 & -0.1354226855 & -22.1795966 & 18.132692 & -4.046905\\
0 & 2 & 0.2651052188 & -0.4010177984 & -0.1359125796 & -22.0338726 & 18.058515 & -3.975358\\
0 & 3 & 0.2620818686 & -0.3987053921 & -0.1366235235 & -21.8180347 & 17.948060 & -3.869975\\
1 & 0 & 0.2463125331 & -0.3772818356 & -0.1309693026 & -20.5906329 & 16.927683 & -3.662950\\
1 & 1 & 0.2453427822 & -0.3765367795 & -0.1311939973 & -20.5213351 & 16.892111 & -3.629225\\
1 & 2 & 0.2434189934 & -0.3750530921 & -0.1316340987 & -20.3837971 & 16.821272 & -3.562525\\
1 & 3 & 0.2405719423 & -0.3728434482 & -0.1322715059 & -20.1800911 & 16.715774 & -3.464317\\
2 & 0 & 0.2255520032 & -0.3521027919 & -0.1265507887 & -19.0111183 & 15.727750 & -3.283368\\
2 & 1 & 0.2246378134 & -0.3513894153 & -0.1267516019 & -18.9456258 & 15.693705 & -3.251921\\
2 & 2 & 0.2228243183 & -0.3499687005 & -0.1271443822 & -18.8156428 & 15.625903 & -3.189740\\
2 & 3 & 0.2201406622 & -0.3478525601 & -0.1277118979 & -18.6231324 & 15.524912 & -3.098220\\
3 & 0 & 0.2057782457 & -0.3277134807 & -0.1219352350 & -17.5054210 & 14.565422 & -2.939999\\
3 & 1 & 0.2049157769 & -0.3270289634 & -0.1221131865 & -17.4434512 & 14.532769 & -2.910683\\
3 & 2 & 0.2032049512 & -0.3256656139 & -0.1224606627 & -17.3204618 & 14.467732 & -2.852729\\
3 & 3 & 0.2006733920 & -0.3236346472 & -0.1229612552 & -17.1383136 & 14.370847 & -2.767466\\
4 & 0 & 0.1868910692 & -0.3040263301 & -0.1171352609 & -16.0656393 & 13.436612 & -2.629027\\
4 & 1 & 0.1860769921 & -0.3033681529 & -0.1172911608 & -16.0069419 & 13.405229 & -2.601713\\
4 & 2 & 0.1844622285 & -0.3020571608 & -0.1175949323 & -15.8904488 & 13.342719 & -2.547730\\
4 & 3 & 0.1820729839 & -0.3001039320 & -0.1180309481 & -15.7179254 & 13.249584 & -2.468342\\
5 & 0 & 0.1688073814 & -0.2809661862 & -0.1121588048 & -14.6849627 & 12.337882 & -2.347080\\
5 & 1 & 0.1680388585 & -0.2803321264 & -0.1122932679 & -14.6293180 & 12.307665 & -2.321653\\
5 & 2 & 0.1665145304 & -0.2790690750 & -0.1125545447 & -14.5188849 & 12.247471 & -2.271414\\
5 & 3 & 0.1642592808 & -0.2771870303 & -0.1129277495 & -14.3553394 & 12.157776 & -2.197563\\
6 & 0 & 0.1514611075 & -0.2584707157 & -0.1070096082 & -13.3576445 & 11.266479 & -2.091165\\
6 & 1 & 0.1507357865 & -0.2578588504 & -0.1071230638 & -13.3048623 & 11.237338 & -2.067524\\
6 & 2 & 0.1492972356 & -0.2566399215 & -0.1073426858 & -13.2001117 & 11.179284 & -2.020828\\
6 & 3 & 0.1471691082 & -0.2548234014 & -0.1076542932 & -13.0449851 & 11.092767 & -1.952218\\
7 & 0 & 0.1348035840 & -0.2364911722 & -0.1016875882 & -12.0790044 & 10.220388 & -1.858616\\
7 & 1 & 0.1341196040 & -0.2358998902 & -0.1017802862 & -12.0289237 & 10.192251 & -1.836673\\
7 & 2 & 0.1327631526 & -0.2347218877 & -0.1019587352 & -11.9295364 & 10.136193 & -1.793344\\
7 & 3 & 0.1307567414 & -0.2329661643 & -0.1022094228 & -11.7823569 & 10.052642 & -1.729715\\
8 & 0 & 0.1188044132 & -0.2149935282 & -0.0961891150 & -10.8454612 & 9.198411 & -1.647050\\
8 & 1 & 0.1181604223 & -0.2144215496 & -0.0962611273 & -10.7979508 & 9.171224 & -1.626727\\
8 & 2 & 0.1168834104 & -0.2132819402 & -0.0963985298 & -10.7036669 & 9.117056 & -1.586611\\
8 & 3 & 0.1149948348 & -0.2115832778 & -0.0965884430 & -10.5640513 & 9.036315 & -1.527737\\
9 & 0 & 0.1034527565 & -0.1939599604 & -0.0905072039 & -9.6545963 & 8.200272 & -1.454324\\
9 & 1 & 0.1028479428 & -0.1934063652 & -0.0905584224 & -9.6095559 & 8.174002 & -1.435554\\
9 & 2 & 0.1016487908 & -0.1923033347 & -0.0906545439 & -9.5201774 & 8.121658 & -1.398519\\
9 & 3 & 0.0998757911 & -0.1906590766 & -0.0907832855 & -9.3878351 & 8.043633 & -1.344202\\
\end{tabular}
\end{ruledtabular}
\end{table*}

\begin{table*}[htb]
\caption{\label{tab:X} 
The electron--proton coupling $A_{ep}=\bar A_{ep} + A^{(6)}_{ep}+A^{(7)}_{ep}$ of Eq.~(\ref{eq:42}) 
and its three parts (in Hz): $A^{(6)}_{ep}$, $A^{(7)}_{ep}$, and $\bar A_{ep}=\langle\pi\delta^{(3)}(r_{ep})\rangle_{vL}\,(1+m/m_p)^{3}\,A^\mathrm{exp}_{p}$, 
with the nonadiabatic (three-body) expectation values $\langle\pi\delta^{(3)}(r_{ep})\rangle_{vL}$ of the first data column 
recalculated here with the CODATA 2022 mass ratios from Table~\ref{tab:VI}.
The numerical uncertainty of $A^{(6)}_{ep}$ is $\sim 5\cdot10^{-5}$ Hz, that of $A^{(7)}_{ep}$ 
is $\sim 2\cdot10^{-4}$ Hz, and that of $\bar A_{ep}$ is $\sim0.01$ Hz (the 12 digits of $\langle\pi\delta^{(3)}\rangle_{vL}$ and the experimental value).
The total uncertainty of $A_{ep}$ (shown in Table~\ref{tab:I}) is dominated by the unknown higher-order terms of Eq.~(\ref{eq:42}).}
\begin{ruledtabular}
\begin{tabular}{mmd{1.12}d{9.2}d{5.5}d{4.4}d{9.2}}
\centt{$v$} & \centt{$L$} & \centt{$\langle\pi\delta^{(3)}(r_{ep})\rangle_{vL}$} & \centt{$\bar A_{ep}$ (Hz)} & \centt{$A^{(6)}_{ep}$ (Hz)} & \centt{$A^{(7)}_{ep}$ (Hz)} & \centt{$A_{ep}$ (Hz)}\\ \hline
0 & 0 & 0.650443509501 & 925404029.05 & -7672.27671 & -538.3160 & 925395818.46\\
0 & 1 & 0.649862629319 & 924577594.70 & -7686.45651 & -533.5476 & 924569374.70\\
0 & 2 & 0.648707155951 & 922933670.68 & -7714.26241 & -524.1148 & 922925432.30\\
0 & 3 & 0.646989479822 & 920489885.19 & -7754.61488 & -510.2210 & 920481620.35\\
1 & 0 & 0.635509211376 & 904156588.72 & -7433.68694 & -482.9266 & 904148672.11\\
1 & 1 & 0.634960524247 & 903375956.32 & -7446.44039 & -478.4803 & 903368031.40\\
1 & 2 & 0.633869164175 & 901823248.06 & -7471.42011 & -469.6865 & 901815306.96\\
1 & 3 & 0.632246981701 & 899515324.05 & -7507.59870 & -456.7387 & 899507359.71\\
2 & 0 & 0.621551029546 & 884297896.13 & -7182.89651 & -432.8822 & 884290280.35\\
2 & 1 & 0.621033039067 & 883560936.70 & -7194.29446 & -428.7362 & 883553313.67\\
2 & 2 & 0.620002813886 & 882095206.75 & -7216.58828 & -420.5382 & 882087569.63\\
2 & 3 & 0.618471693396 & 879916839.15 & -7248.79990 & -408.4721 & 879909181.88\\
3 & 0 & 0.608513974390 & 865749716.02 & -6920.92228 & -387.6121 & 865742407.48\\
3 & 1 & 0.608025345054 & 865054529.50 & -6931.02263 & -383.7470 & 865047214.73\\
3 & 2 & 0.607053597129 & 863671996.77 & -6950.74503 & -376.1065 & 863664669.92\\
3 & 3 & 0.605609585591 & 861617561.48 & -6979.15816 & -364.8653 & 861610217.46\\
4 & 0 & 0.596348750375 & 848441914.26 & -6648.48054 & -346.6134 & 848434919.16\\
4 & 1 & 0.595888295787 & 847786812.74 & -6657.32926 & -343.0122 & 847779812.40\\
4 & 2 & 0.594972664819 & 846484119.16 & -6674.57103 & -335.8951 & 846477108.69\\
4 & 3 & 0.593612252966 & 844548623.46 & -6699.31885 & -325.4284 & 844541598.72\\
5 & 0 & 0.585011187222 & 832311648.57 & -6366.02186 & -309.4412 & 832304973.11\\
5 & 1 & 0.584577859909 & 831695141.78 & -6373.65385 & -306.0888 & 831688462.04\\
5 & 2 & 0.583716262653 & 830469323.46 & -6388.48366 & -299.4652 & 830462635.51\\
5 & 3 & 0.582436354837 & 828648363.09 & -6409.66639 & -289.7287 & 828641663.70\\
6 & 0 & 0.574461716426 & 817302623.75 & -6073.75859 & -275.7011 & 817296274.29\\
6 & 1 & 0.574054598849 & 816723406.28 & -6080.19822 & -272.5843 & 816717053.50\\
6 & 2 & 0.573245211232 & 815571868.06 & -6092.66375 & -266.4278 & 815565508.97\\
6 & 3 & 0.572043098825 & 813861589.38 & -6110.35027 & -257.3822 & 813855221.65\\
7 & 0 & 0.564664866368 & 803364373.34 & -5771.68605 & -245.0417 & 803358356.61\\
7 & 1 & 0.564283162614 & 802821312.82 & -5776.94749 & -242.1486 & 802815293.72\\
7 & 2 & 0.563524403186 & 801741804.72 & -5787.07608 & -236.4361 & 801735781.21\\
7 & 3 & 0.562397739942 & 800138869.67 & -5801.30486 & -228.0472 & 800132840.31\\
8 & 0 & 0.555588737742 & 790451513.30 & -5459.59820 & -217.1486 & 790445836.56\\
8 & 1 & 0.555231765090 & 789943638.41 & -5463.68555 & -214.4692 & 789937960.25\\
8 & 2 & 0.554522278195 & 788934231.71 & -5471.48437 & -209.1803 & 788928551.05\\
8 & 3 & 0.553469055070 & 787435781.95 & -5482.26365 & -201.4182 & 787430098.27\\
9 & 0 & 0.547204397912 & 778522880.38 & -5137.09859 & -191.7394 & 778517551.55\\
9 & 1 & 0.546871576857 & 778049366.63 & -5140.00571 & -189.2648 & 778044037.36\\
9 & 2 & 0.546210212619 & 777108425.38 & -5145.46146 & -184.3821 & 777103095.54\\
9 & 3 & 0.545228727647 & 775712039.48 & -5152.76871 & -177.2209 & 775706709.49\\
\end{tabular}
\end{ruledtabular}
\end{table*}

\begin{table*}[htb]
\caption{\label{tab:XI} The electron--deuteron coupling $A_{ed}=\bar A_{ed} + A^{(6)}_{ed}+A^{(7)}_{ed}$ of Eq.~(\ref{eq:42}) 
and its three parts (in Hz): $A^{(6)}_{ed}$, $A^{(7)}_{ed}$, and $\bar A_{ed}=\langle\pi\delta^{(3)}(r_{ed})\rangle_{vL}\,(1+m/m_d)^{3}\,A^\mathrm{exp}_{d}$, 
with the nonadiabatic (three-body) expectation values $\langle\pi\delta^{(3)}(r_{ed})\rangle_{vL}$ of the first data column 
recalculated here with the CODATA 2022 mass ratios from Table~\ref{tab:VI}.
The numerical uncertainty of $A^{(6)}_{ed}$ is $\sim 10^{-5}$ Hz, that of $A^{(7)}_{ed}$ 
is $\sim 3\cdot10^{-5}$ Hz, and that of $\bar A_{ed}$ is $\sim0.01$ Hz (the 12 digits of $\langle\pi\delta^{(3)}\rangle_{vL}$ and the experimental value).
The total uncertainty of $A_{ed}$ (shown in Table~\ref{tab:I}) is dominated by the unknown higher-order terms of Eq.~(\ref{eq:42}).}

\begin{ruledtabular}
\begin{tabular}{mmd{1.12}d{9.2}d{5.5}d{3.4}d{9.2}}
\centt{$v$} & \centt{$L$} & \centt{$\langle\pi\delta^{(3)}(r_{ed})\rangle_{vL}$} & \centt{$\bar A_{ed}$ (Hz)} & \centt{$A^{(6)}_{ed}$ (Hz)} & \centt{$A^{(7)}_{ed}$ (Hz)} & \centt{$A_{ed}$ (Hz)}\\ \hline
0 & 0 & 0.651403398955 & 142289087.29 & -1177.74131 & -82.6348 & 142287826.92\\
0 & 1 & 0.650822518150 & 142162202.78 & -1179.91800 & -81.9028 & 142160940.96\\
0 & 2 & 0.649667048722 & 141909808.20 & -1184.18637 & -80.4548 & 141908543.56\\
0 & 3 & 0.647949391364 & 141534612.28 & -1190.38072 & -78.3220 & 141533343.58\\
1 & 0 & 0.636490374341 & 139031565.66 & -1141.11632 & -74.1322 & 139030350.42\\
1 & 1 & 0.635941784883 & 138911734.71 & -1143.07406 & -73.4496 & 138910518.18\\
1 & 2 & 0.634850625540 & 138673387.67 & -1146.90860 & -72.0998 & 138672168.67\\
1 & 3 & 0.633228757533 & 138319115.47 & -1152.46223 & -70.1122 & 138317892.89\\
2 & 0 & 0.622559347762 & 135988546.45 & -1102.61846 & -66.4501 & 135987377.38\\
2 & 1 & 0.622041569564 & 135875445.74 & -1104.36812 & -65.8136 & 135874275.56\\
2 & 2 & 0.621011774696 & 135650502.84 & -1107.79036 & -64.5552 & 135649330.49\\
2 & 3 & 0.619481313982 & 135316197.16 & -1112.73504 & -62.7030 & 135315021.72\\
3 & 0 & 0.609556441086 & 133148260.81 & -1062.40382 & -59.5008 & 133147138.90\\
3 & 1 & 0.609068160321 & 133041603.36 & -1063.95428 & -58.9075 & 133040480.50\\
3 & 2 & 0.608097115890 & 132829493.59 & -1066.98179 & -57.7346 & 132828368.87\\
3 & 3 & 0.606654175408 & 132514305.36 & -1071.34338 & -56.0090 & 132513178.01\\
4 & 0 & 0.597433818898 & 130500259.82 & -1020.58235 & -53.2073 & 130499186.03\\
4 & 1 & 0.596973878228 & 130399792.83 & -1021.94068 & -52.6545 & 130398718.23\\
4 & 2 & 0.596059282375 & 130200013.38 & -1024.58740 & -51.5619 & 130198937.23\\
4 & 3 & 0.594700441367 & 129903195.39 & -1028.38634 & -49.9553 & 129902117.05\\
5 & 0 & 0.586149251644 & 128035319.08 & -977.22322 & -47.5011 & 128034294.35\\
5 & 1 & 0.585716642996 & 127940822.35 & -978.39477 & -46.9865 & 127939796.97\\
5 & 2 & 0.584856491769 & 127752935.50 & -980.67124 & -45.9697 & 127751908.86\\
5 & 3 & 0.583578774760 & 127473837.80 & -983.92292 & -44.4751 & 127472809.41\\
6 & 0 & 0.575665787676 & 125745367.07 & -932.35902 & -42.3218 & 125744392.39\\
6 & 1 & 0.575259647743 & 125656652.03 & -933.34754 & -41.8433 & 125655676.84\\
6 & 2 & 0.574452226273 & 125480283.19 & -935.26108 & -40.8983 & 125479307.03\\
6 & 3 & 0.573253090307 & 125218350.32 & -937.97607 & -39.5097 & 125217372.83\\
7 & 0 & 0.565951544832 & 123623439.63 & -885.98904 & -37.6154 & 123622516.03\\
7 & 1 & 0.565571153722 & 123540349.02 & -886.79671 & -37.1713 & 123539425.06\\
7 & 2 & 0.564815033739 & 123375186.20 & -888.35151 & -36.2944 & 123374261.56\\
7 & 3 & 0.563692365217 & 123129956.47 & -890.53571 & -35.0066 & 123129030.93\\
8 & 0 & 0.556979647512 & 121663666.19 & -838.08165 & -33.3336 & 121662794.77\\
8 & 1 & 0.556624431394 & 121586074.67 & -838.70908 & -32.9223 & 121585203.04\\
8 & 2 & 0.555918476824 & 121431869.72 & -839.90625 & -32.1105 & 121430997.70\\
8 & 3 & 0.554870600663 & 121202977.23 & -841.56094 & -30.9189 & 121202104.75\\
9 & 0 & 0.548728361954 & 119861299.34 & -788.57599 & -29.4332 & 119860481.33\\
9 & 1 & 0.548397901339 & 119789115.29 & -789.02225 & -29.0533 & 119788297.21\\
9 & 2 & 0.547741285305 & 119645687.58 & -789.85975 & -28.3038 & 119644869.41\\
9 & 3 & 0.546766991555 & 119432868.04 & -790.98145 & -27.2045 & 119432049.86\\
\end{tabular}
\end{ruledtabular}
\end{table*}

\end{document}